\documentclass[letterpaper,journal]{IEEEtran}
\usepackage{amsmath,amsfonts}
\usepackage{amssymb}
\usepackage{amsthm}
\usepackage{cite}
\usepackage{footnote}
\usepackage{stfloats}
\usepackage{array}
\usepackage{enumerate}
\usepackage{multirow}
\usepackage{balance}
\usepackage{algorithm}
\usepackage{algpseudocode}
\usepackage{graphicx}
\usepackage{subfigure}
\usepackage{textcomp}
\usepackage{color}
\usepackage{xcolor}
\usepackage{tabularx}
\usepackage{booktabs}
\usepackage{epsfig}
\usepackage{verbatim}
\usepackage{hyperref}
\usepackage{makecell}

\newcommand{\cmark}{\ensuremath{\checkmark}}
\newcommand{\xmark}{\ensuremath{\times}}

\theoremstyle{definition}
\newtheorem{theorem}{Theorem}

\begin{document}


\title{Covert Semantic Transmission in ISAC: Dual-Functional Waveform Design and \\ Rectified Flow-Assisted Recovery}

\author{Yunfan~Bai,~\IEEEmembership{Graduate Student Member,~IEEE}, Yuwen~Qian, Cheng~Zeng,~\IEEEmembership{Member,~IEEE}, Zhen~Mei,~\IEEEmembership{Member,~IEEE}, Zhaohui~Yang,~\IEEEmembership{Member,~IEEE}, Wei~Zhu, Shuning~Zhang, Feng~Shu,~\IEEEmembership{Senior Member,~IEEE}, and Jun~Li,~\IEEEmembership{Fellow,~IEEE}

\thanks{This work was supported in part by the National Natural Science Foundation of China under Grant 62671315.}
\thanks{This work has been submitted to the IEEE for possible publication. Copyright may be transferred without notice, after which this version may no longer be accessible.}

\thanks{Yunfan~Bai, Yuwen~Qian, Cheng~Zeng, Zhen~Mei, and Shuning~Zhang are with the School of Electronic and Optical Engineering, Nanjing University of Science and Technology, Nanjing 210094, China (e-mail:\{baiyunfan, admon, czeng, meizhen, and zhangshuning\}@njust.edu.cn).}

\thanks{Zhaohui Yang is with the College of Information Science and Electronic Engineering, Zhejiang University, Hangzhou, Zhejiang 310027, China (e-mail:
yang\_zhaohui@zju.edu.cn).}

\thanks{Wei~Zhu is with the School of Energy and Power Engineering, Nanjing University of Science and Technology, Nanjing 210094, China. (e-mail: zhuwei@njust.edu.cn).}

\thanks{Feng~Shu is with the School of Information and Communication Engineering and Collaborative Innovation Center of Information Technology, Hainan University, Haikou 570228, China, and also with the School of Electronic and Optical Engineering, Nanjing University of Science and Technology, Nanjing 210094, China. (e-mail: shufeng0101@163.com).}

\thanks{Jun~Li is with the School of Information Science and Engineering, Southeast University, Nanjing 210096, China. (e-mail: jun.li@seu.edu.cn).}

}

\maketitle

\begin{abstract}
Semantic integrated sensing and communication~(ISAC) is envisioned as a promising paradigm for efficient and intelligent connectivity in future wireless networks.
However, the open wireless channel exposes the dual-functional waveform to detection, which challenges the joint guarantee of covertness, sensing fidelity, and semantic accuracy.
To address the challenge, we propose CoSMIC, a novel covertness-oriented semantic ISAC framework, where the sensing output is embedded into a dual-functional ISAC waveform through semantic modulation.
Specifically, a semantic rotation coding scheme is established to map semantic latents onto the pairwise rotation and scaling of Gaussian reference sequences, which satisfies a derived closed-form covertness constraint by a differentiable budget projection.
Moreover, the radar performance is analyzed to confirm an invariant matched-filter mainlobe response and a bounded output signal-to-interference-plus-noise ratio~(SINR) under the semantic embedding.
Subsequently, a reliability-guided rectified flow~(RFlow) refiner is designed to effectively reconstruct high-fidelity semantic representations from coarse observations.
Simulation results demonstrate that CoSMIC improves the semantic reconstruction quality by 18\% over diffusion-based baseline schemes with substantially reduced inference latency under strict covertness constraints, which validates the applicability to practical ISAC scenarios.
The source code and video demonstrations are available at \url{https://github.com/LanceAnlan/CoSMIC-covertness-oriented-semantic-ISAC-framework}.

\end{abstract}


\vspace{-1em}
\section{Introduction}

\IEEEPARstart{T}{he} sixth-generation (6G) wireless networks are envisioned to provide intelligent connectivity and environmental awareness for emerging applications, including autonomous driving and low-altitude Internet of Things~(IoT) networks.
To address the requirement for intelligent connectivity, semantic communication~(SemCom) conveys the underlying meaning of the source instead of the exact bits, and improves the communication efficiency under constrained bandwidth~\cite{xie2021deep}.
Concurrently, integrated sensing and communication (ISAC) provides a waveform-level foundation for joint environmental perception and data transmission, which further supports semantic ISAC by conveying the scene and target semantics extracted from each sensing frame to the receiver~\cite{yang2025integrated}.
\looseness=-1

Driven by deep learning~(DL), SemCom has progressed from image reconstruction toward task-oriented and multimodal transmission, where scene, state, and linguistic semantics are jointly conveyed \cite{park2025transmit,jiang2025visual,weng2026generative}.
While the conveyed content grows richer, the semantics delivered over the open wireless channels expose private information regarding the user and the environment, which raises pronounced security and privacy concerns~\cite{won2025resource}.
The exposure becomes more acute in semantic ISAC, where the dual-functional waveform openly carries semantic information within reach of a passive adversary.
In sensitive applications, including surveillance and autonomous driving, the leakage of scene and target semantics threatens to expose location and behavior.
Conventional encryption protects the message content from disclosure, whereas the existence of the link stays exposed to a vigilant adversary~\cite{guo2025survey}.
A more fundamental safeguard conceals the presence of the link rather than the content exclusively, which prevents the warden from detecting the ongoing communication.
Therefore, concealing the semantics carried by the ISAC waveform constitutes the core security challenge, which remains unaddressed in existing efficiency-oriented designs.

Recent research on semantic ISAC has pursued the joint design of sensing and semantic transmission, and the transmitted content has evolved from raw observations to task-relevant semantics.
To exploit the integration between perception and communication, a semantic-driven framework was developed to fuse multimodal sensing into compact transmittable semantics~\cite{peng2026simac}.
A large artificial intelligence model was further developed to fuse heterogeneous sensing modalities adaptively for multimodal ISAC~\cite{peng2026lam}.
Within a unified ISAC design, digital semantic transmission and radar sensing were jointly optimized to balance the reconstruction quality and the sensing accuracy~\cite{huan2026joint}.
For low-altitude intelligent networks, an intelligent semantic scheme was designed to extract and deliver semantics under sensing guidance~\cite{liu2026intelligent}.
For edge video transmission, computer vision sensing was employed to identify the dynamic content of the scene and to adapt the semantic transmission accordingly~\cite{peng2026semantic}.
In multi-device video analytics, a task-oriented architecture conveyed only the semantics required by the downstream inference \cite{he2026task}.
At the network level, beamforming and resource management were jointly optimized for ISAC networking with multiple unmanned aerial vehicles~(UAVs)~\cite{zhou2026beamforming}.
At the waveform level, a MIMO-OFDM ISAC waveform was shaped to suppress the range-Doppler sidelobes and preserve the sensing performance~\cite{li2025mimo}.
However, the above designs focus on improving transmission efficiency and sensing accuracy based on openly radiated waveforms, without considering the security of the conveyed semantic information. 
The lack of covertness makes private semantic data vulnerable to passive eavesdropping, which is a critical unsolved problem in semantic ISAC systems.


In contrast to conventional security methods that encrypt the message content, covert communication establishes an orthogonal dimension of security by concealing the existence of the transmission.
The fundamental limit of covert communication is captured by the square-root law, which bounds the information delivered while remaining undetected by the warden~\cite{bash2013limits}.
In classical intra-pulse radar-embedded communication, a tag remodulated each incident radar pulse into a communication waveform concealed within the ambient radar backscatter~\cite{blunt2010intrapulse}, and the performance of the intended and intercept receivers was characterized through dedicated metrics~\cite{blunt2011performance}.
Given that the sensing waveform is inherently radiated, it serves as a natural vehicle for data embedding, which motivates the paradigm of covert ISAC~\cite{hu2024covertisac}.
Cooperative NOMA and reconfigurable intelligent surfaces were further exploited to improve the covertness~\cite{li2025covert,zhao2026sensing}.
Against multiple or colluding wardens, robust beamforming kept the detection error probability high enough to preserve covertness~\cite{wu2025covertisac}.
Meanwhile, full-duplex jamming was employed to confuse the warden during simultaneous sensing and communication on a UAV platform~\cite{wang2026uav}.
Through the cooperation between a radar and a communication transmitter, the radar probing signal was also exploited as the interference toward the warden~\cite{wang2024sensing}.
Under imperfect channel knowledge, a robust transceiver sustained covertness despite the residual uncertainty~\cite{zhang2025robust}.
In a similar setting, dual-functional artificial noise with a varying power was transmitted to confuse the warden and to assist the target sensing~\cite{tang2025dual}.
Nevertheless, the covert ISAC schemes above conceal a separate communication signal through beamforming, artificial noise, or cooperative jamming, which requires additional spatial degrees of freedom or transmit power together with the channel or location information of the warden that is usually unavailable in practice.
Beyond the implementation barrier, the covert message remains at the bit level, which leaves high-dimensional scene semantics outside the scope of covert protection.
\looseness=-1

To convey meaning under covertness, covert SemCom transmits compact semantics and masks sensitive images with covertness-oriented artificial noise~\cite{yin2026semantic}.
A MIMO spatial modulation scheme further conveys covert semantics through the antenna index, where an excess distortion exponent bound guides the training of the SemCom network~\cite{bai2026mimo}.
Adaptive dual-path encoding and private semantic optimization were introduced to balance recovery quality and covertness~\cite{zhang2026optimization}.
Since covert transmission operates at low power under artificial noise, the legitimate channel degrades severely and demands stronger recovery at the receiver.
To satisfy the rigorous requirement, diffusion models were placed at the receiver to denoise the semantic features and restore the content~\cite{wu2024cddm,guo2025diffusion}.
A diffusion-based time-frequency framework was also developed to improve the signal quality in bistatic ISAC~\cite{wang2026tfdiffusion}.
In parallel, exploiting a shared semantic knowledge base allows the receiver to regenerate the content from a highly compact representation~\cite{li2025semantic}.
To further optimize the transmission efficiency, the spatio-temporal variation in semantic reliability has motivated unequal protection of the conveyed content~\cite{tung2025multi}.
Although diffusion-based refiners improve recovery quality, their iterative sampling process incurs extremely high inference latency, which conflicts with the low-latency requirement of real-time ISAC systems. 
Flow matching~\cite{lipman2023flow} and rectified flow~\cite{liu2023flow} instead learn a velocity field that transports a source distribution to the target distribution along straight paths, which supports generation within a few integration steps.
More fundamentally, there is still a research gap in achieving covertness, semantic efficiency, and sensing fidelity simultaneously in a unified ISAC framework.

To address these issues, we propose a novel covert semantic ISAC framework. 
In particular, the main contributions of this paper are listed as follows:

\begin{itemize}
    \item We propose CoSMIC, a covert semantic ISAC system where a dual-functional waveform senses the environment and conveys the resulting semantics to the receiver in the next frame. Furthermore, a Gaussian pair rotation coding scheme is designed to embed the covert semantics in the phase of the sensing waveform through the rotation of reference Gaussian pairs.

    \item To guarantee covertness against an arbitrary warden, a closed-form lower bound on the optimal detection error probability is derived via the Kullback--Leibler~(KL) divergence at the transmitted waveform, where the bound holds without the location information of the warden. A differentiable covert projection is then formulated to restrict each transmitted frame within the covertness constraint.
\looseness=-1

    \item The invariance of the matched-filter mainlobe under semantic embedding is established, and a lower bound on the radar output SINR is derived to validate the preservation of sensing performance.

    \item 
    A reliability-guided rectified flow~(RFlow) refiner is developed to counteract the signal-dependent nonlinear distortions in CoSMIC. By leveraging a continuous transport trajectory guided by reliability features, the RFlow refiner effectively maps coarse observations back to the clean semantic manifold with significantly reduced latency.

 \end{itemize}


The remainder of this paper is organized as follows.
Section~\ref{sec: system_model} presents the system model of the proposed CoSMIC.
Section~\ref{sec:covert_design} derives the covertness constraint, develops the covert projection, and analyzes the radar performance.
Section~\ref{sec:network} describes the CoSMIC network with the RFlow-assisted receiver.
Section~\ref{sec:results} provides the simulation results, and Section~\ref{sec:conclusion} concludes this paper.

Throughout the paper, boldface lowercase and boldface uppercase letters denote vectors and matrices, respectively.
The operators $(\cdot)^{\mathsf{T}}$, $(\cdot)^{\mathsf{H}}$, and $\mathbb{E}[\cdot]$ represent the transpose, the conjugate transpose, and the expectation, respectively.
$\mathcal{CN}(\boldsymbol{\mu},\boldsymbol{\Sigma})$ stands for the circularly symmetric complex Gaussian distribution with mean $\boldsymbol{\mu}$ and covariance $\boldsymbol{\Sigma}$, while $\mathcal{N}(\boldsymbol{\mu},\boldsymbol{\Sigma})$ stands for the real Gaussian counterpart.
In addition, $\mathbf{I}_N$ denotes the $N\times N$ identity matrix, and $D(\cdot\|\cdot)$ denotes the KL divergence.
The main variables of this paper are summarized in Table~\ref{tab:notation}.

\begin{table}[t]
\centering
\caption{Summary of Main Variables}
\label{tab:notation}
\vspace{-0.4em}
\renewcommand{\arraystretch}{1.0}
\setlength{\tabcolsep}{4pt}
\begin{tabular}{lp{6.0cm}}
\hline
\textbf{Symbol} & \textbf{Definition} \\
\hline
$M$, $N$, $J$ & Numbers of chirps, samples per chirp, and pairs \\
$\mathbf{m}_k$, $\mathbf{i}_k$, $\boldsymbol{\varsigma}_k$ & Semantic message with image and state components \\
$r_{k,\ell}$, $v_{k,\ell}$, $\theta_{k,\ell}$ & Range, velocity, and angle of the $\ell$-th target \\
$B$, $T_c$, $T_s$ & Bandwidth, chirp duration, and sampling interval \\
$p_n$, $\mu$ & Discrete chirp sample and normalized chirp rate \\
$x_{k,m,n}^{(i)}$, $\phi_{k,m,n}^{(i)}$ & Transmitted sample and phase under mode $i$ \\
$H_0$, $H_1$ & Sensing-only and covert SemCom hypotheses \\
$\xi_k$, $\epsilon$ & Detection error probability and covertness budget \\
$\mathbf{q}_k$, $\mathbf{b}_k^{(i)}$ & Reference and mode-$i$ symbol vectors \\
$\mathbf{A}_k$, $\tau_{k,j}$ & Rotation matrix and scale of the $j$-th pair \\
$\rho$, $\alpha_{\phi}$ & Shaping coefficient and phase mapping gain \\
$\lambda_k$ & Contraction coefficient of the covert projection \\
$\Delta$, $\Gamma$, $\eta$ & Range lag, Doppler lag, and confidence level \\
$E_{\Theta}$, $\mathbf{v}_{\Psi}$, $D_{\Omega}$ & Semantic encoder, velocity field, and decoder \\
\hline
\end{tabular}
\vspace{-0.8em}
\end{table}

\section{System Model}\label{sec: system_model}
\vspace{-0.25em}

\begin{figure*}[htb]
    \centering
    \includegraphics[width=0.9\textwidth]{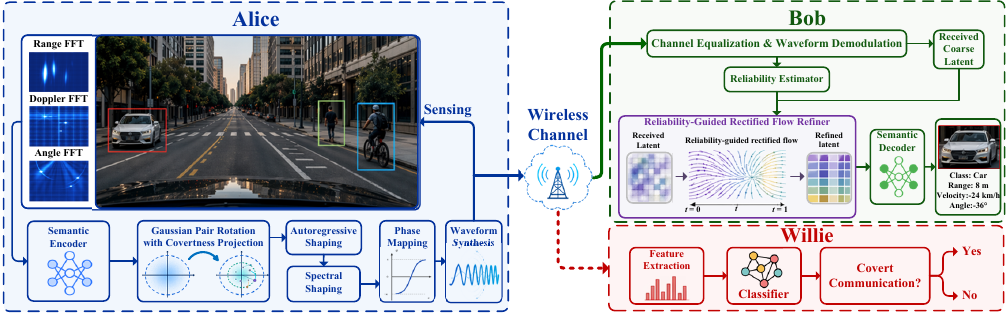}
\vspace{-0.5em}
    \caption{System model of the proposed CoSMIC framework, consisting of a dual-functional transmitter (Alice), a legitimate receiver (Bob), and a passive warden (Willie) with DL-based detectors.
}
    \label{system_model}
\vspace{-0.8em}
\end{figure*}

As shown in Fig.~\ref{system_model}, we propose a frame-based covert semantic ISAC system~(CoSMIC).
In each frame, Alice transmits an ISAC waveform to probe the sensing region and simultaneously conveys covert semantic information to Bob.
Then, Alice extracts the target information from the received echoes and combines the estimates with the scene image as the semantic message for the next frame.
At the legitimate receiver, Bob demodulates the waveform to obtain the image and target states.
Meanwhile, Willie monitors the wireless environment with a binary classifier to detect whether the covert communication exists.

\subsection{CoSMIC Framework}\label{subsec:cosmic_framework}
\vspace{-0.25em}

The proposed CoSMIC operates on a frame basis, where Alice transmits with a single antenna and collects target echoes with a uniform linear array of $U$ elements, while Bob and Willie are each equipped with a single antenna.
The $k$-th frame contains $M$ chirps, and each chirp is sampled into $N$ baseband points.
Since each unit-modulus sample contributes one controllable real phase, one frame provides $MN$ real degrees of freedom for semantic embedding.

The sensing and communication functions are linked through a framewise feedback loop.
Let $\mathbf{o}_{k-1}$ denote the sensing output of the $(k-1)$-th frame, from which Alice forms the semantic message at the beginning of the $k$-th frame as
\begin{equation}\label{eq:semantic_payload}
\mathbf{m}_k=\Phi\left(\mathbf{o}_{k-1}\right)=\left[\mathbf{i}_k,\boldsymbol{\varsigma}_k\right],
\end{equation}
where $\mathbf{i}_k$ denotes the image semantic component, $\boldsymbol{\varsigma}_k=\{(r_{k,\ell},v_{k,\ell},\theta_{k,\ell},g_{k,\ell})\}_{\ell}$ stacks the target range, velocity, angle, and class label, and $\Phi(\cdot)$ maps the sensing output to the current semantic message.
This framewise embedding at the waveform level differs from the sensing-guided adaptive transmission in~\cite{peng2026semantic} and complements the joint beamforming and resource optimization in~\cite{zhou2026beamforming}.

The covert communication link and the monitoring link are modeled as quasi-static Rayleigh fading channels over each chirp.
Let $h_{k,m}^{b}$ and $h_{k,m}^{w}$ denote the channel coefficients from Alice to Bob and from Alice to Willie during the $m$-th chirp of the $k$-th frame, respectively, and let $\mathbf{n}_{k,m}^{b}$ and $\mathbf{n}_{k,m}^{w}$ denote the corresponding additive white Gaussian noise (AWGN) vectors.
For $u\in\{b,w\}$, the channel and noise terms satisfy
\begin{equation}\label{eq:channel_statistics}
h_{k,m}^{u}\sim\mathcal{CN}(0,1),\quad
\mathbf{n}_{k,m}^{u}\sim\mathcal{CN}\left(\mathbf{0},\sigma_u^{2}\mathbf{I}_N\right),
\end{equation}
where $\sigma_b^{2}$ and $\sigma_w^{2}$ denote the noise powers at Bob and Willie, respectively.

\subsection{Dual-Functional Signal Model}\label{subsec:signal_model}
\vspace{-0.25em}

CoSMIC adopts the chirp signal as the shared carrier of the sensing and communication functions.
The continuous-time form of a single chirp is expressed as
\begin{equation}\label{eq:chirp_continuous}
s_c(t)=\exp\left(j2\pi\left(f_c t+\frac{B}{2T_c}t^{2}\right)\right),\quad 0\le t<T_c,
\end{equation}
where $f_c$ denotes the carrier frequency, $B$ denotes the sweep bandwidth, and $T_c$ denotes the chirp duration.
After baseband sampling at $t=nT_s$ with the fast-time sampling interval $T_s=T_c/N$, the quadratic phase $\pi(B/T_c)t^{2}$ becomes $\pi BT_sn^{2}/N$.
Accordingly, the $n$-th sample of the discrete chirp is written as
\begin{equation}\label{eq:chirp_discrete}
p_n=\exp\left(j\pi\mu\frac{n^{2}}{N}\right),\quad n=0,1,\ldots,N-1,
\end{equation}
where $\mu=BT_s$ denotes the normalized chirp rate that equals the ratio of the sweep bandwidth to the sampling rate $1/T_s$, and $|p_n|=1$ holds for all $n$.

Based on the chirp, information embedding is achieved by phase rotation of unit-modulus samples.
The $n$-th transmitted sample of the $m$-th chirp in the $k$-th frame is given by
\begin{equation}\label{eq:transmit_signal}
x_{k,m,n}^{(i)}=p_n e^{j\phi_{k,m,n}^{(i)}},\quad i\in\{0,1\},
\end{equation}
where $\phi_{k,m,n}^{(i)}$ denotes the modulated phase, and the superscript $i$ indicates the transmit mode.
Specifically, $i=0$ corresponds to the sensing-only mode driven by a reference random sequence, while $i=1$ represents the covert SemCom mode.
Since $|p_n|=1$, the relation $|x_{k,m,n}^{(i)}|=1$ holds under both modes, which keeps a constant envelope and identical radiated power across the sensing-only and covert SemCom modes.
Therefore, the semantic embedding avoids an explicit average-power signature at Willie.

For the communication function, let $\mathbf{x}_{k,m}^{(i)}=[x_{k,m,0}^{(i)},\ldots,x_{k,m,N-1}^{(i)}]^{\mathsf{T}}\in\mathbb{C}^{N}$ collect the transmitted samples of the $m$-th chirp.
When the semantic waveform is transmitted, the received signal at Bob is expressed as
\begin{equation}\label{eq:bob_receive}
\mathbf{y}_{k,m}^{b}=h_{k,m}^{b}\mathbf{x}_{k,m}^{(1)}+\mathbf{n}_{k,m}^{b},
\end{equation}
where $\mathbf{y}_{k,m}^{b}\in\mathbb{C}^{N}$ denotes the received vector of the $m$-th chirp in the $k$-th frame, and $h_{k,m}^{b}$ and $\mathbf{n}_{k,m}^{b}$ follow the statistics in~\eqref{eq:channel_statistics}.

For the sensing function, Alice collects the target echoes with the uniform linear array of $U$ elements, where the steering vector is defined as
\begin{equation}\label{eq:steering_vector}
\mathbf{a}(\theta)=\left[1,\ e^{-j\psi(\theta)},\ \ldots,\ e^{-j(U-1)\psi(\theta)}\right]^{\mathsf{T}},
\end{equation}
where $\psi(\theta)=\frac{2\pi d_a}{\lambda_c}\sin\theta$, $\theta$ is the target angle, $d_a$ is the inter-element spacing, $\lambda_c=c_0/f_c$ is the carrier wavelength ($c_0$ being the speed of light), and $\|\mathbf{a}(\theta)\|_2^{2}=U$.
Suppose that $L_k$ targets are present in the $k$-th frame, where the $\ell$-th target is characterized by the parameter tuple $\boldsymbol{\zeta}_{k,\ell}=(d_{k,\ell},\nu_{k,\ell},\theta_{k,\ell},\alpha_{k,\ell})$.
Specifically, $d_{k,\ell}$, $\nu_{k,\ell}$, $\theta_{k,\ell}$, and $\alpha_{k,\ell}\in\mathbb{C}$ denote the range bin, the Doppler bin, the angle, and the complex reflection coefficient, respectively.
The echo received at the $a$-th array element is then expressed as
\begin{equation}\label{eq:radar_echo}
y_{k,m,n,a}^{r}=\sum_{\ell=1}^{L_k}\alpha_{k,\ell}a_a(\theta_{k,\ell})\omega_{k,\ell,m}x_{k,m,n-d_{k,\ell}}^{(i)}+w_{k,m,n,a}^{r},
\end{equation}
where $\omega_{k,\ell,m}=e^{j2\pi\nu_{k,\ell}m/M}$ denotes the Doppler phasor, $a_a(\theta_{k,\ell})$ denotes the $a$-th entry of $\mathbf{a}(\theta_{k,\ell})$, and $w_{k,m,n,a}^{r}\sim\mathcal{CN}(0,\sigma_r^{2})$ denotes the radar noise.

To extract the target parameters, Alice performs matched filtering over the range, Doppler, and angle domains, where the three-dimensional matched response is defined as
\begin{equation}\label{eq:matched_response}
R_k(d,\nu,\theta)=\sum_{a=0}^{U-1}a_a^{*}(\theta)\!\sum_{m=0}^{M-1}\!\sum_{n=0}^{N-1}y_{k,m,n,a}^{r}x_{k,m,n-d}^{(i)*}e^{-j2\pi\nu m/M}.
\end{equation}
The parameter estimate of the $\ell$-th target is obtained from the local peak of the matched response as
\begin{equation}\label{eq:parameter_estimation}
(\hat{d}_{k,\ell},\hat{\nu}_{k,\ell},\hat{\theta}_{k,\ell})=\arg\max_{(d,\nu,\theta)\in\mathcal{P}_{\ell}}\left|R_k(d,\nu,\theta)\right|^{2},
\end{equation}
where $\mathcal{P}_{\ell}$ denotes the search region around the $\ell$-th local peak.
The discrete estimates are further converted into physical parameters as
\begin{equation}\label{eq:meta_conversion}
\hat{r}_{k,\ell}=\frac{c_0 T_s}{2}\hat{d}_{k,\ell},\quad
\hat{v}_{k,\ell}=\frac{\lambda_c}{2MT_r}\hat{\nu}_{k,\ell},
\end{equation}
where $T_r$ denotes the chirp repetition interval, and the angle estimate $\hat{\theta}_{k,\ell}$ is adopted directly.
The class estimate $\hat{g}_{k,\ell}$ of the $\ell$-th target is identified by Alice from the scene image at the position of the target.
Together with the class estimates $\{\hat{g}_{k,\ell}\}_{\ell=1}^{L_k}$ and the scene image, the physical estimates form $\mathbf{o}_k$ and generate $\mathbf{m}_{k+1}$ through~\eqref{eq:semantic_payload}.


\subsection{Binary Hypothesis Testing at Willie}\label{subsec:willie_detection}
\vspace{-0.25em}

It is assumed that Willie monitors the wireless channel to determine whether semantic transmission is hidden in the sensing waveform.
The received signal at Willie under the two transmit modes is given by
\begin{equation}\label{eq:willie_receive}
\mathbf{y}_{k,m}^{w}=h_{k,m}^{w}\mathbf{x}_{k,m}^{(i)}+\mathbf{n}_{k,m}^{w},\quad i\in\{0,1\}.
\end{equation}
Since the fading coefficient $h_{k,m}^{w}$ is unknown to Willie, the observation conditioned on a fixed transmitted chirp follows
\begin{equation}\label{eq:willie_conditional}
\mathbf{y}_{k,m}^{w}\mid\mathbf{x}_{k,m}^{(i)}\sim\mathcal{CN}\left(\mathbf{0},\ \sigma_w^{2}\mathbf{I}_N+\mathbf{x}_{k,m}^{(i)}\mathbf{x}_{k,m}^{(i)\mathsf{H}}\right),
\end{equation}
where the rank-one covariance term results from the marginalization over the Rayleigh fading.

Let $\mathbf{y}_k^{w}=\{\mathbf{y}_{k,m}^{w}\}_{m=0}^{M-1}$ denote the whole-frame observation at Willie, and let $W_k(\cdot\mid\mathbf{x})$ denote the channel transition law in~\eqref{eq:willie_conditional}.
The frame-level marginal distribution under mode $i$ is given by
\begin{equation}\label{eq:frame_marginal}
P_{i,k}=\int W_k(\cdot\mid\mathbf{x})\,dP_{\mathbf{x},i,k}(\mathbf{x}),\quad i\in\{0,1\},
\end{equation}
where $P_{\mathbf{x},i,k}$ denotes the distribution of the transmitted waveform in the $k$-th frame.
Accordingly, Willie faces the binary hypothesis testing problem
\begin{equation}\label{eq:hypothesis_test}
H_0:\ \mathbf{y}_k^{w}\sim P_{0,k},\qquad
H_1:\ \mathbf{y}_k^{w}\sim P_{1,k},
\end{equation}
where $H_0$ corresponds to the sensing-only mode and $H_1$ corresponds to the covert SemCom mode.

In practical scenarios, Willie is assumed to extract statistical features from $\mathbf{y}_k^{w}$ and adopt a DL-based classifier to distinguish $H_0$ from $H_1$.
The DL-based classifier captures high-order distributional differences beyond the average power, which poses a more severe threat to the covert transmission~\cite{qian2025adversarial}.
The detection performance of Willie is measured by the total error probability $\xi_k=P_{f,k}+P_{m,k}$, where $P_{f,k}$ and $P_{m,k}$ denote the false alarm and missed detection probabilities, respectively.
For any decision rule adopted by Willie, including a DL classifier, the detection error probability $\xi_k$ is lower bounded by that of the optimal likelihood ratio test.
Accordingly, the covertness analysis in Section~\ref{sec:covert_design} is performed under the optimal test, and the derived guarantee applies to general detector structures.


\subsection{Gaussian Pair Rotation Coding}\label{subsec:rotation_coding}
\vspace{-0.25em}

The phase sequences of CoSMIC are generated from a real-valued symbol vector that is organized in two-dimensional pairs.
Since each frame provides $MN$ real degrees of freedom, the symbol vector contains $J=MN/2$ pairs.
For the $k$-th frame, each reference pair follows
\begin{equation}\label{eq:reference_field}
\mathbf{q}_{k,j}\sim\mathcal{N}(\mathbf{0},\mathbf{I}_2),\quad j=1,\ldots,J,
\end{equation}
and stacked into the reference symbol vector $\mathbf{q}_k=[\mathbf{q}_{k,1}^{\mathsf{T}},\ldots,\mathbf{q}_{k,J}^{\mathsf{T}}]^{\mathsf{T}}\in\mathbb{R}^{MN}$, where the pairs are expanded frame by frame from a secret seed shared between Alice and Bob.
Each frame thus uses fresh reference pairs, and the shared seed is renewed secretly at irregular intervals.
The pairs are independent of the semantic message, since the semantic encoder determines the rotation parameters without access to the reference pairs.

To embed semantics, the semantic encoder maps the message $\mathbf{m}_k$ to pair-level parameters $(c_{k,j},s_{k,j})$, which define the rotation matrix
\begin{equation}\label{eq:rotation_matrix}
\mathbf{A}_{k,j}=\begin{bmatrix}c_{k,j} & -s_{k,j}\\s_{k,j} & c_{k,j}\end{bmatrix},
\mathbf{A}_k=\mathrm{blkdiag}\left(\mathbf{A}_{k,1},\ldots,\mathbf{A}_{k,J}\right).
\end{equation}
Geometrically, $\mathbf{A}_{k,j}$ rotates the $j$-th reference pair and scales the pair energy by $\tau_{k,j}=c_{k,j}^{2}+s_{k,j}^{2}$.
As a result, the semantic message is carried by the rotation angles and the scale factors of the $J$ pairs.

Accordingly, the symbol vectors under the sensing-only mode and the covert SemCom mode are expressed as
\begin{equation}\label{eq:frontend_fields}
\mathbf{b}_k^{(0)}=\mathbf{q}_k,\quad
\mathbf{b}_k^{(1)}=\mathbf{A}_k\mathbf{q}_k,
\end{equation}
where the sensing-only mode uses the reference vector, and the covert SemCom mode uses the corresponding rotated vector.
Before the chirp modulation, the symbol vectors under both modes pass through the same shaping operation as
\begin{equation}\label{eq:ar_shaping}
\mathbf{u}_k^{(i)}=\mathbf{G}_{\rho}\mathbf{b}_k^{(i)},\quad i\in\{0,1\},
\end{equation}
where $\mathbf{G}_{\rho}\in\mathbb{R}^{MN\times MN}$ denotes an invertible lower triangular matrix that imposes first-order autoregressive correlation with coefficient $\rho$, $|\rho|<1$.
The shaped vector is then partitioned into $M$ segments of length $N$, and the entry that corresponds to the $n$-th sample of the $m$-th chirp is mapped to the phase
\begin{equation}\label{eq:phase_mapping}
\phi_{k,m,n}^{(i)}=\pi\tanh\!\left(\alpha_{\phi} u_{k,m,n}^{(i)}\right),
\end{equation}
where $\alpha_{\phi}>0$ denotes the phase mapping gain, and the mapping confines the phase within $(-\pi,\pi)$ in a strictly monotonic manner.
The substitution of~\eqref{eq:phase_mapping} into~\eqref{eq:transmit_signal} completes the waveform generation.
Therefore, the sensing-only mode and the covert SemCom mode share the chirp basis, the shaping matrix, and the phase mapping.
The difference between the transmit modes lies in the symbol vector, which forms the basis of the covertness analysis.


\section{Covertness-Constrained Waveform Design and Radar Performance Analysis}\label{sec:covert_design}
\vspace{-0.25em}

In the proposed CoSMIC, the pair parameters $(c_{k,j},s_{k,j})$ in Section~\ref{subsec:rotation_coding} jointly determine the detectability at Willie and the matched response at Alice.
Accordingly, this section first derives a closed-form covertness constraint on the pair scales, then develops the covert projection for the waveform generation, and finally analyzes the radar performance of the generated waveform.
\vspace{-1em}

\subsection{Covertness Constraint}\label{subsec:covertness_constraint}
\vspace{-0.25em}

Let $\mathbf{y}^{w,(K)}=\{\mathbf{y}_k^{w}\}_{k=1}^{K}$ denote the observation of Willie over a window of $K$ consecutive frames, and let $P_{0}^{(K)}$ and $P_{1}^{(K)}$ denote the joint distributions of $\mathbf{y}^{w,(K)}$ under $H_0$ and $H_1$.
The total error probability of Willie's detection based on $\mathbf{y}^{w,(K)}$ is denoted by $\xi^{(K)}$.
For a prescribed covertness budget $\epsilon\in(0,1)$, the covertness requirement is formulated as $\xi^{(K)}\ge 1-\epsilon$, which indicates that Willie's optimal detection error probability is close to that of random guessing, and the covert transmission cannot be effectively detected.
According to the covert communication theory in~\cite{bash2013limits}, the optimal detection error probability satisfies
\begin{equation}\label{eq:pinsker_covertness}
\xi^{(K)}\ge 1-\sqrt{\frac{D\left(P_{1}^{(K)}\,\middle\|\,P_{0}^{(K)}\right)}{2}}.
\end{equation}
Therefore, the KL divergence between $P_{1}^{(K)}$ and $P_{0}^{(K)}$ provides a tractable covertness measure, where a sufficient condition for satisfying $\xi^{(K)}\ge 1-\epsilon$ is expressed as
\begin{equation}\label{eq:kl_covertness_constraint}
D\left(P_{1}^{(K)}\,\middle\|\,P_{0}^{(K)}\right)\le 2\epsilon^{2}.
\end{equation}

To characterize the divergence, we first define
\begin{equation}\label{eq:chi_function}
\chi(t)=t-\ln t-1,\quad t>0,
\end{equation}
which is convex and attains the unique minimum $\chi(1)=0$.
Then, the closed-form covertness constraint is given in Theorem~\ref{thm:covertness}.

\begin{theorem}[\textbf{Multi-Frame Covertness Constraint}]\label{thm:covertness}
For an arbitrary window of $K$ consecutive frames with the transmit waveforms generated by \eqref{eq:frontend_fields}--\eqref{eq:phase_mapping} and \eqref{eq:transmit_signal}, the joint distributions $P_{0}^{(K)}$ and $P_{1}^{(K)}$ of the observations at Willie over the window satisfy
\begin{equation}\label{eq:kl_bound}
D\left(P_{1}^{(K)}\,\middle\|\,P_{0}^{(K)}\right)\le\mathbb{E}\left[\sum_{k=1}^{K}\sum_{j=1}^{J}\chi(\tau_{k,j})\right].
\end{equation}
To guarantee the covertness requirement $\xi^{(K)}\ge 1-\epsilon$ at Willie over the window, it suffices that the pair scales satisfy
\begin{equation}\label{eq:covert_budget}
\sum_{k=1}^{K}\sum_{j=1}^{J}\chi(\tau_{k,j})\le 2\epsilon^{2}.
\end{equation}
\end{theorem}

\IEEEproof 
See Appendix~\ref{app:proof_theorem1}. 
\endIEEEproof



Theorem~\ref{thm:covertness} provides a design rule for covert semantic embedding by separating the semantic rotation from the covertness cost.
The bound holds for an arbitrary detector without knowledge of the seed, which makes the guarantee information-theoretic.
The rotation angles are absent from the KL-divergence bound and therefore carry semantic information without consuming the covertness budget.
In contrast, the radial scales $\tau_{k,j}$ enter the bound through the convex cost $\chi(\tau_{k,j})$, which allocates the budget $2\epsilon^{2}$ to energy scaling and vanishes at $\tau_{k,j}=1$.
The budget is allocated evenly over a monitoring window of $K_{\mathrm{W}}$ frames, where each frame is assigned the share $2\epsilon^{2}/K_{\mathrm{W}}$ and the accumulated KL divergence over the window stays within $2\epsilon^{2}$.
The separation supports rich semantic modulation under a strict covertness budget, and the enforcement of~\eqref{eq:covert_budget} during waveform generation is presented in the next subsection.


\begin{figure} 
  \centering
  \includegraphics[width=\linewidth]{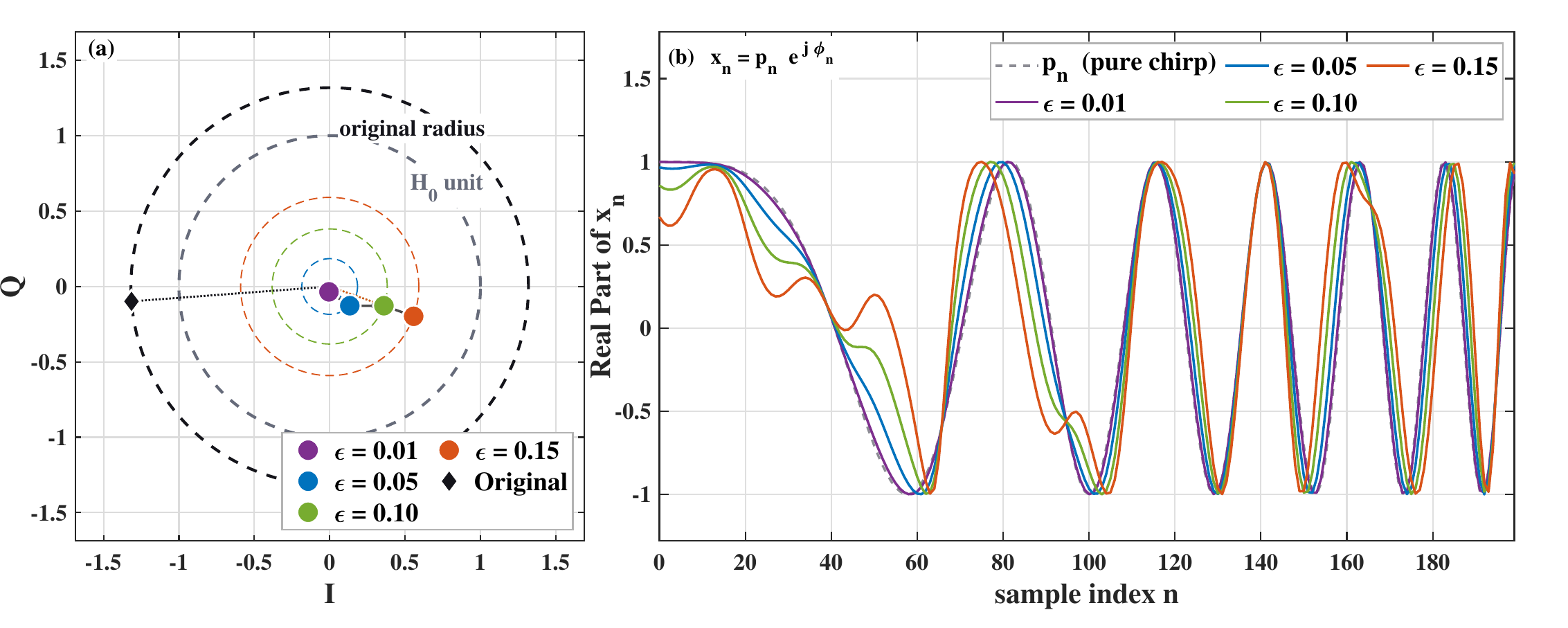}
\vspace{-0.4em}
  \caption{Illustration of semantic-guided pair rotation.}
  \label{gaussian_rotation}
\vspace{-0.6em}
\end{figure}

\subsection{Covert Projection and Waveform Generation}\label{subsec:covert_projection}
\vspace{-0.25em}

The covertness constraint in Theorem~\ref{thm:covertness} is enforced during waveform generation through a per-frame projection on the pair scales.
For the message $\mathbf{m}_k$, the semantic encoder $E_{\Theta}(\cdot)$ with parameters $\Theta$ first produces the raw pair
\begin{equation}\label{eq:raw_pair}
(\tilde{c}_{k,j},\tilde{s}_{k,j})=E_{\Theta}(\mathbf{m}_k),\quad j=1,\ldots,J.
\end{equation}
Each raw pair is decomposed into
\begin{equation}\label{eq:radius_direction}
\tilde{\tau}_{k,j}=\tilde{c}_{k,j}^{2}+\tilde{s}_{k,j}^{2},\quad
\bar{c}_{k,j}=\frac{\tilde{c}_{k,j}}{\sqrt{\tilde{\tau}_{k,j}}},\quad
\bar{s}_{k,j}=\frac{\tilde{s}_{k,j}}{\sqrt{\tilde{\tau}_{k,j}}},
\end{equation}
where the normalized pair $(\bar{c}_{k,j},\bar{s}_{k,j})$ determines the rotation angle of the $j$-th pair, and $\tilde{\tau}_{k,j}$ represents the energy level assigned by the encoder.
Following Theorem~\ref{thm:covertness}, the covertness budget is governed by the scale terms, and the projection preserves the normalized directions while contracting the scales toward unity.
To this end, let $\lambda\in[0,1]$ denote the scale retention coefficient between unit scales and raw encoder scales. 
The budget function is defined as
\begin{equation}\label{eq:budget_function}
F_k(\lambda)=\sum_{j=1}^{J}\chi\left(1+\lambda(\tilde{\tau}_{k,j}-1)\right).
\end{equation}
Then, the contraction coefficient is obtained as
\begin{equation}\label{eq:contraction}
\lambda_k=\sup\left\{\lambda\in[0,1]:F_k(\lambda)\le 2\epsilon^{2}/K_{\mathrm{W}}\right\}.
\end{equation}
Since $\chi(\cdot)$ is convex with $\chi(1)=0$, the budget function $F_k(\lambda)$ is continuous and nondecreasing on $[0,1]$ with $F_k(0)=0$.
Therefore, the contraction coefficient $\lambda_k$ is uniquely determined and equals unity when the requested transmission satisfies the budget.
Accordingly, the projected scales and the transmitted pair parameters are expressed as
\begin{equation}\label{eq:projected_pair}
\tau_{k,j}=1+\lambda_k(\tilde{\tau}_{k,j}-1),
c_{k,j}=\sqrt{\tau_{k,j}}\,\bar{c}_{k,j},
s_{k,j}=\sqrt{\tau_{k,j}}\,\bar{s}_{k,j},
\end{equation}
which satisfy 
\begin{equation}
c_{k,j}^{2}+s_{k,j}^{2}=\tau_{k,j},\sum_{j=1}^{J}\chi(\tau_{k,j})=F_k(\lambda_k)\le 2\epsilon^{2}/K_{\mathrm{W}}.
\end{equation}
Thus, each transmitted frame satisfies the prescribed covertness budget by construction.
The projected parameters are used to form the rotation matrix $\mathbf{A}_{k,j}$ in~\eqref{eq:rotation_matrix}, and the rotated field $\mathbf{b}_k^{(1)}=\mathbf{A}_k\mathbf{q}_k$ is further shaped and modulated through~\eqref{eq:transmit_signal} to generate the transmit waveform.

Fig.~\ref{gaussian_rotation} illustrates how the scale projection modifies the Gaussian reference pairs and the resulting transmit signal.
In panel~(a), each reference pair is rotated along the encoded direction and projected toward the unit circle, where a smaller covertness budget $\epsilon$ yields a tighter contraction. 
In panel~(b), the real part of the corresponding transmit signal remains close to the pure chirp, and the deviation grows with $\epsilon$.
\vspace{-2em}


\begin{figure*}[b]

\vspace{-0.3em}
\hrulefill
\vspace{-0.4em}
\begin{equation}\label{eq:envelopes}
B_{A,k}(\Delta,\Gamma;\eta)=\left|\bar{A}_k^{(0)}(\Delta,\Gamma)\right|+\delta_{\tau}+2L_1\sqrt{\ln(4/\eta)},\quad
B_{E,k}(\Delta,\Gamma;\eta)=\delta_{\tau}+2(L_0+L_1)\sqrt{\ln(4/\eta)}.
\end{equation}

\vspace{-0.9em}
\begin{equation}\label{eq:sinr_bound}
\mathrm{SINR}_{k,\ell}^{(1)}=\frac{|\alpha_{k,\ell}UMN|^{2}}{|I_{k,\ell}^{(1)}|^{2}+\sigma_r^{2}UMN}\ge\frac{|\alpha_{k,\ell}|^{2}(UMN)^{2}}{\left(\sum_{r\ne\ell}|\alpha_{k,r}|\,|S(\theta_{k,\ell},\theta_{k,r})|\,B_{A,k}\!\left(\Delta_{\ell r},\Gamma_{\ell r};\tfrac{\eta}{R_k^{\mathrm{I}}}\right)\right)^{2}+\sigma_r^{2}UMN}.
\end{equation}

\vspace{-0.7em}
\end{figure*}

\subsection{Radar Performance Analysis}\label{subsec:radar_analysis}
\vspace{-0.25em}

The semantic rotation modifies the phase of the transmit waveform and thus alters the matched response.
With the cyclic fast-time index $\langle n-\Delta\rangle_N=(n-\Delta)\bmod N$, the cyclic ambiguity function under mode $i$ is defined as
\begin{equation}\label{eq:ambiguity}
A_k^{(i)}(\Delta,\Gamma)=\sum_{m=0}^{M-1}\sum_{n=0}^{N-1}x_{k,m,n}^{(i)}x_{k,m,\langle n-\Delta\rangle_N}^{(i)*}e^{-j2\pi\Gamma m/M},
\end{equation}
where $\Delta$ and $\Gamma$ denote the range and Doppler lags.
With the array inner product $S(\theta,\vartheta)=\mathbf{a}^{\mathsf{H}}(\theta)\mathbf{a}(\vartheta)$ and $S(\theta,\theta)=U$, the matched response in \eqref{eq:matched_response} under the mode-$i$ waveform becomes
\begin{equation}\label{eq:matched_mode}
R_k^{(i)}(d,\nu,\theta)=\sum_{\ell=1}^{L_k}\alpha_{k,\ell}S(\theta,\theta_{k,\ell})A_k^{(i)}(d-d_{k,\ell},\nu-\nu_{k,\ell})+Z_k^{(i)},
\end{equation}
where $Z_k^{(i)}\sim\mathcal{CN}(0,\sigma_r^{2}UMN)$ denotes the filtered noise, and $R_{k,s}^{(i)}$ denotes the noise-free signal component.

The sidelobe behavior is governed by the mean term $\bar{A}_k^{(0)}(\Delta,\Gamma)=\mathbb{E}[A_k^{(0)}(\Delta,\Gamma)]$, which equals $\kappa_0^{2}r_p(\Delta)D_M(\Gamma)$ for $\rho=0$ with the mean phasor $\kappa_0=\mathbb{E}[e^{j\phi_{k,m,n}^{(0)}}]$, the chirp autocorrelation $r_p(\Delta)=\sum_{n=0}^{N-1}p_np_{\langle n-\Delta\rangle_N}^{*}$, and the Doppler sum $D_M(\Gamma)=\sum_{m=0}^{M-1}e^{-j2\pi\Gamma m/M}$.
At a nonzero lag, $A_k^{(i)}(\Delta,\Gamma)$ is a Lipschitz function of the Gaussian reference vector $\mathbf{q}_k$ with the constant $L_i$ under $H_i$, where $L_0=2\pi\alpha_{\phi}\sqrt{MN}\,\|\mathbf{G}_{\rho}\|_{2}$ and $L_1=L_0\sqrt{\tau_{\max}}$.
Herein, $\|\mathbf{G}_{\rho}\|_{2}\le\sqrt{(1+|\rho|)/(1-|\rho|)}$ holds, and $\tau_{\max}$ denotes the largest scale admitted by the budget with $\chi(\tau_{\max})=2\epsilon^{2}$.
The scales shift the mean term by at most $\delta_{\tau}=2\epsilon L_0\sqrt{h_{\max}}$, where $h_{\max}$ denotes the maximum of $(\sqrt{\tau}-1)^{2}/\chi(\tau)$ over the admitted scales.
For any $\eta\in(0,1)$, the Gaussian concentration of Lipschitz functions keeps the sidelobe magnitude under $H_1$ and the inter-mode sidelobe deviation within the envelopes $B_{A,k}(\Delta,\Gamma;\eta)$ and $B_{E,k}(\Delta,\Gamma;\eta)$ in~\eqref{eq:envelopes} with probability at least $1-\eta$, which holds for every rotation field and every scale map within the covertness budget.
The impact of the semantic rotation coding on the radar performance is then given in Theorem~\ref{thm:radar}.

\begin{theorem}[\textbf{Mainlobe Invariance and Multi-Target SINR Bound}]
\label{thm:radar}
Consider an arbitrary frame $k$ in which the targets are characterized by $\{\boldsymbol{\zeta}_{k,\ell}\}_{\ell=1}^{L_k}$.

\emph{\textbf{Single-target case:}} If $L_k=1$, the noise-free response at the true bin is invariant to the transmit mode and satisfies
\begin{equation}\label{eq:mainlobe_invariance}
R_{k,s}^{(0)}(d_{k,1},\nu_{k,1},\theta_{k,1})=R_{k,s}^{(1)}(d_{k,1},\nu_{k,1},\theta_{k,1})=\alpha_{k,1} UMN,
\end{equation}
which yields $\mathrm{SNR}_k^{(0)}=\mathrm{SNR}_k^{(1)}=|\alpha_{k,1}|^{2}UMN/\sigma_r^{2}$ for the output signal-to-noise ratio~(SNR).

\emph{\textbf{Multi-target case:}} If $L_k\ge 2$, let $R_k^{\mathrm{I}}=L_k-1$, $\Delta_{\ell r}=d_{k,\ell}-d_{k,r}$, and $\Gamma_{\ell r}=\nu_{k,\ell}-\nu_{k,r}$.
With probability at least $1-\eta$, the response deviation at each true bin is bounded by
\begin{equation}\label{eq:response_dev}
\left|R_{k,s}^{(1)}-R_{k,s}^{(0)}\right|\le\sum_{r\ne\ell}\!|\alpha_{k,r}||S(\theta_{k,\ell},\theta_{k,r})|B_{E,k}\!\left(\Delta_{\ell r},\Gamma_{\ell r};\tfrac{\eta}{R_k^{\mathrm{I}}}\right),
\end{equation}
and the output signal-to-interference-plus-noise ratio~(SINR) of the $\ell$-th target satisfies the lower bound in \eqref{eq:sinr_bound}.
\end{theorem}

\IEEEproof
See Appendix~\ref{app:proof_theorem2}.
\endIEEEproof


Theorem~\ref{thm:radar} indicates that the mainlobe and the output SNR of an isolated target are unaffected by the semantic embedding.
In multi-target scenes, the sidelobe deviation and the residual interference are bounded by $B_{E,k}$ and $B_{A,k}$, which keeps the sensing accuracy predictable.
Therefore, CoSMIC supports semantic communication over the dual-functional sensing waveform without requiring an additional radar-side compensation stage to preserve the sensing functionality.

\vspace{-0.25em}


\section{CoSMIC Network}\label{sec:network}
\vspace{-0.2em}

\begin{figure*}[htb]
    \centering
    \includegraphics[width=0.95\textwidth]{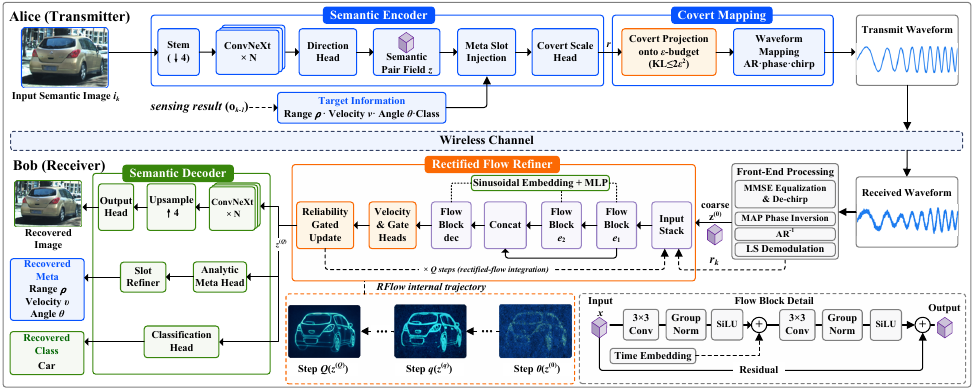}
\vspace{-0.5em}
    \caption{Network architecture of the proposed CoSMIC, where the transmitter implements the semantic encoder and covert mapping, while the receiver incorporates the reliability-guided RFlow refiner to decode the semantic results.
}
    \label{network}
\vspace{-0.8em}
\end{figure*}

The complete architecture of the CoSMIC network is depicted in Fig.~\ref{network}, where the transmitter maps the semantic message into the pair field and the receiver recovers the semantic message from the covert waveform.
At Bob, the received signal is first equalized, and a coarse semantic latent is then demodulated through a sequence of signal processing operations.
To compensate for the residual distortion, an RFlow refiner guided by the reliability features transports the coarse latent toward the clean latent.
Subsequently, a semantic decoder restores the scene image and the target states.
\vspace{-1em}

\subsection{Transmitter Network}\label{subsec:tx_network}
\vspace{-0.25em}

As illustrated in Fig.~\ref{network}, the semantic information $\mathbf{m}_k=[\mathbf{i}_k,\boldsymbol{\varsigma}_k]$ is mapped by the transmitter network into a pair of parameters that modulate the covert waveform.
The transmitter network implements the semantic encoder $E_{\Theta}$ of \eqref{eq:raw_pair}, where a shared backbone drives a direction head and a covert scale head to generate the directions and raw scales for the $J$ pairs.
The image component $\mathbf{i}_k\in\mathbb{R}^{H_{\mathrm{i}}\times W_{\mathrm{i}}\times 3}$ is first passed through a stem layer, where a strided convolution reduces the spatial resolution and increases the channel dimension.
The resulting feature map is refined by $N_{\mathrm{c}}$ cascaded ConvNeXt blocks~\cite{liu2022convnet}.
Each block consists of a depthwise convolution, layer normalization, and a pointwise feedforward unit with a residual connection.
The hierarchical visual feature is then obtained as
\begin{equation}\label{eq:convnext}
\mathbf{F}_k=\mathrm{ConvNeXt}^{N_{\mathrm{c}}}\!\left(\mathrm{Stem}(\mathbf{i}_k)\right)\in\mathbb{R}^{H\times W\times C},
\end{equation}
where $H$, $W$, and $C$ denote the reduced height, reduced width, and channel dimension, respectively.

To incorporate the sensing context, the state component $\boldsymbol{\varsigma}_k$ represents the range, velocity, angle, and class of the targets estimated in the previous frame.
The state variables are then embedded as meta tokens and injected into the flattened visual tokens through the meta slot injection module.
Let $\mathcal{A}$ denote the index set of the state components, and let $\mathcal{J}_{p,\ell}$ denote the reserved pair positions for component $p\in\mathcal{A}$ of the $\ell$-th target.
The feature map $\mathbf{F}_k$ is projected to $J$ pair tokens, and the positions outside $\bigcup_{p,\ell}\mathcal{J}_{p,\ell}$ carry the image semantics.
Each component of the $\ell$-th target is normalized to $\bar{\varsigma}_{k,p,\ell}\in[-1,1]$ and mapped to a rotation angle $\omega_{k,p,\ell}=\gamma_{p}\bar{\varsigma}_{k,p,\ell}$ with a fixed gain $\gamma_p$, and the meta slot injection embeds the state onto the reserved positions as
\begin{equation}\label{eq:meta_slot_injection}
(\bar{c}_{k,j},\bar{s}_{k,j})=(\cos\omega_{k,p,\ell},\sin\omega_{k,p,\ell}),\quad \tilde{\tau}_{k,j}=1,
\end{equation}
where $j\in\mathcal{J}_{p,\ell}$, $p\in\mathcal{A}$, and $\ell=1,\ldots,L_k$.
The angle is replicated over $\mathcal{J}_{p,\ell}$ for redundant protection against the channel noise, and the unit scale keeps the reserved pairs free of covertness cost.
Bob recovers each component through a reliability-weighted circular mean of the received angles over $\mathcal{J}_{p,\ell}$, which yields a gain of $10\log_{10}|\mathcal{J}_{p,\ell}|$~dB.
The size of $\mathcal{J}_{p,\ell}$ is therefore set by the precision required for each component.

For each pair token outside the reserved positions, a pair of heads derives the parameters of the corresponding Gaussian pair.
The direction head outputs a two-dimensional vector, which is normalized to unit norm and used as the semantic pair field, and the normalized direction specifies the rotation angle of the $j$-th pair.
In parallel, the covert scale head predicts a raw scale through a softplus activation to ensure positivity, yielding
\begin{equation}
\begin{gathered}\label{eq:tx_heads}
(\bar{c}_{k,j},\bar{s}_{k,j})=\mathrm{DirHead}(\mathbf{z}_{k,j}),\\
 \tilde{\tau}_{k,j}=\mathrm{softplus}\!\left(\mathrm{ScaleHead}(\mathbf{z}_{k,j})\right),
\end{gathered}
\end{equation}
where $\bar{c}_{k,j}^{2}+\bar{s}_{k,j}^{2}=1$, and $(\bar{c}_{k,j},\bar{s}_{k,j})$ corresponds to the encoder output in~\eqref{eq:radius_direction}.
The raw scale $\tilde{\tau}_{k,j}$ is adjusted by the covert projection in~\eqref{eq:contraction}--\eqref{eq:projected_pair} to satisfy the covertness budget.
The projected parameters then form the rotation matrix in~\eqref{eq:rotation_matrix}.
The covert projection and waveform mapping constitute smooth mappings of the pair parameters, allowing the gradient to propagate through the differentiable chain to $E_{\Theta}$ during training.
Specifically, the contraction coefficient $\lambda_k$ is computed by bisection on $F_k(\lambda)$ in the forward pass and treated as a constant in the backward pass, which yields $\partial\tau_{k,j}/\partial\tilde{\tau}_{k,j}=\lambda_k$ from~\eqref{eq:projected_pair}.
After the autoregressive shaping in~\eqref{eq:ar_shaping}, the phase mapping in~\eqref{eq:phase_mapping}, and the chirp modulation in~\eqref{eq:transmit_signal}, the dual-functional transmit waveform is generated to probe the targets and convey the covert semantics to Bob.

\subsection{RFlow-Assisted Receiver Network}
\vspace{-0.25em}

At Bob, each chirp is first equalized by the minimum mean square error (MMSE) filter
\begin{equation}\label{eq:mmse}
\hat{\mathbf{x}}_{k,m}=\frac{h_{k,m}^{b*}}{|h_{k,m}^{b}|^{2}+\sigma_b^{2}}\,\mathbf{y}_{k,m}^{b},
\end{equation}
and the chirp carrier is removed to expose the modulated phase $\hat{\phi}_{k,m,n}=\angle(\hat{x}_{k,m,n}p_n^{*})$.
The phase map in~\eqref{eq:phase_mapping} is inverted by the maximum a posteriori (MAP) estimate
\begin{equation}\label{eq:map}
\hat{u}_{k,m,n}=\arg\min_{u}\left[\frac{(\hat{\phi}_{k,m,n}-\pi\tanh(\alpha_{\phi}u))^{2}}{2\sigma_{\phi,k,m,n}^{2}}+\frac{u^{2}}{2}\right],
\end{equation}
where $\sigma_{\phi,k,m,n}^{2}$ denotes the phase noise variance of the sample.
The autoregressive shaping is then reversed through $\hat{\mathbf{b}}_k=\mathbf{G}_{\rho}^{-1}\hat{\mathbf{u}}_k$, and a pairwise least squares projection on the reference pair $\mathbf{q}_{k,j}$ returns the coarse estimates $(\hat{c}_{k,j},\hat{s}_{k,j})$.
With the pair magnitude $\hat{m}_{k,j}=\sqrt{\hat{c}_{k,j}^{2}+\hat{s}_{k,j}^{2}}$, the coarse semantic latent and the reliability features are arranged as
\begin{equation}\label{eq:coarse_latent}
\begin{gathered}
\mathbf{z}_k^{(0)}=\mathcal{M}\!\left\{\frac{\hat{c}_{k,j}}{\hat{m}_{k,j}},\frac{\hat{s}_{k,j}}{\hat{m}_{k,j}}\right\}_{j=1}^{J}, \\
\mathbf{r}_k=\mathcal{M}\!\left\{\log e_{k,j},\log\hat{m}_{k,j},\log\gamma_{k,j},\delta_{k,j}\right\}_{j=1}^{J},
\end{gathered}
\end{equation}
where $\mathcal{M}$ reorders the pair quantities into the latent map, $e_{k,j}$ denotes the reference-pair energy, $\gamma_{k,j}$ denotes the receive SNR proxy, and $\delta_{k,j}$ denotes the phase residual.
 

Unlike the artificial Gaussian corruption assumed by diffusion denoisers, the coarse latent distortion in CoSMIC stems from fading equalization, nonlinear phase-map inversion, and covert scale contraction.
Such distortion forms a signal-dependent and directionally constrained residual, whereas a diffusion-style isotropic score correction risks disturbing the angular pair structure that carries the semantics.
To address this issue, an RFlow refiner is designed to transport the coarse latent toward the clean latent along a continuous trajectory under the guidance of the reliability features.

Let $\boldsymbol{\theta}_k^{\star}$ collect the pair angles of the clean latent formed by the transmitter-side directions $(\bar{c}_{k,j},\bar{s}_{k,j})$ used as supervision during training, and let $\boldsymbol{\theta}_k^{(0)}$ collect the pair angles of the coarse latent in \eqref{eq:coarse_latent}.
Since every pair direction lies on the unit circle, the latent of the $J$ pairs lives on the torus $\mathbb{T}^{J}=(\mathcal{S}^{1})^{J}$, where the logarithm map and the exponential map of the circle read $\mathrm{Log}_{a}(b)=\mathrm{w}(b-a)$ and $\mathrm{Exp}_{a}(v)=\mathrm{w}(a+v)$ with $\mathrm{w}(\cdot)$ wrapping an angle into $[-\pi,\pi)$.
The refiner transports $\boldsymbol{\theta}_k^{(0)}$ toward $\boldsymbol{\theta}_k^{\star}$ along the geodesic bridge
\begin{equation}\label{eq:bridge}
\boldsymbol{\theta}_{k,t}=\mathrm{Exp}_{\boldsymbol{\theta}_k^{(0)}}\!\left(t\,\boldsymbol{\Delta}_k\right),\quad \boldsymbol{\Delta}_k=\mathrm{Log}_{\boldsymbol{\theta}_k^{(0)}}\!\left(\boldsymbol{\theta}_k^{\star}\right),
\end{equation}
which is the shortest path on $\mathbb{T}^{J}$ for $t\in[0,1]$ and moves with the constant velocity $\boldsymbol{\Delta}_k$.
Then, a velocity field network $\mathbf{v}_{\Psi}(\cdot)$ with parameters $\Psi$ predicts the instantaneous flow drift conditioned on the bridge state, time, and reliability features, optimized via the flow matching risk
\begin{equation}\label{eq:reflow_risk}
\mathcal{L}_f(\Psi)=\int_{0}^{1}\mathbb{E}\!\left[\left\|\boldsymbol{\Delta}_k-\mathbf{v}_{\Psi}(\boldsymbol{\theta}_{k,t},t,\mathbf{r}_k)\right\|_2^{2}\right]dt.
\end{equation}
The reliability features enter the velocity field through the input channels, a feature-wise affine modulation of every block, and a per-pair gate on the predicted velocity.
The risk is minimized by the conditional expectation of the angular residual, which yields the optimal velocity field
\begin{equation}\label{eq:reflow_optimal}
\mathbf{v}^{\star}(\boldsymbol{\theta},t,\mathbf{r})=\mathbb{E}\!\left[\boldsymbol{\Delta}_k\,\middle|\,\boldsymbol{\theta}_{k,t}=\boldsymbol{\theta},\,\mathbf{r}_k=\mathbf{r}\right].
\end{equation}
Substitution of~\eqref{eq:reflow_optimal} into~\eqref{eq:reflow_risk} reduces the optimal risk to the residual covariance, and the law of total covariance yields
\begin{equation}\label{eq:reflow_cov}
\begin{aligned}
\mathcal{L}_f^{\star}
&=\int_{0}^{1}\mathbb{E}\,\mathrm{tr}\,\mathrm{Cov}\!\left(\boldsymbol{\Delta}_k\,\middle|\,\boldsymbol{\theta}_{k,t},\mathbf{r}_k\right)dt\\
&\le\int_{0}^{1}\mathbb{E}\,\mathrm{tr}\,\mathrm{Cov}\!\left(\boldsymbol{\Delta}_k\,\middle|\,\boldsymbol{\theta}_{k,t}\right)dt,
\end{aligned}
\end{equation}
which shows that the reliability conditioning does not increase the attainable flow matching risk, with a strict reduction whenever $\mathbf{r}_k$ is informative about the residual.

At inference, the trained velocity field drives the refinement through the ordinary differential equation
\begin{equation}\label{eq:reflow_ode}
\frac{d\boldsymbol{\theta}_{k,t}}{dt}=\mathbf{v}_{\Psi}(\boldsymbol{\theta}_{k,t},t,\mathbf{r}_k),\quad \boldsymbol{\theta}_{k,0}=\boldsymbol{\theta}_k^{(0)},
\end{equation}
which starts at the coarse latent and follows the learned trajectory toward the clean latent.
The trajectory is discretized into $Q$ explicit steps on the torus as
\begin{equation}\label{eq:reflow_step}
\boldsymbol{\theta}_k^{(q+1)}=\mathrm{Exp}_{\boldsymbol{\theta}_k^{(q)}}\!\left(\frac{1}{Q}\,\mathbf{v}_{\Psi}\!\left(\boldsymbol{\theta}_k^{(q)},\frac{q}{Q},\mathbf{r}_k\right)\right),
\end{equation}
where $q=0,\ldots,Q-1$, the step size $1/Q$ advances the virtual time, and $\mathbf{v}_k^{(q)}$ denotes the velocity predicted at the $q$-th step.
The exponential map rotates every pair as $e^{j\theta_{k,j}^{(q+1)}}=e^{j\theta_{k,j}^{(q)}}e^{jv_{k,j}^{(q)}/Q}$ and keeps the unit modulus of the pair direction.
Driven by the conditional velocity $\boldsymbol{\Delta}_k$, the update reproduces the bridge in~\eqref{eq:bridge} for every $Q$, and the training and the inference therefore share the same state space and the same update.
To align the velocity field with the states visited by the discrete update, the training also unrolls~\eqref{eq:reflow_step} and minimizes the rollout consistency risk
\begin{equation}\label{eq:reflow_roll}
\begin{aligned}
\mathcal{L}_r(\Psi)&=\frac{1}{Q}\sum_{q=0}^{Q-1}\mathbb{E}\!\left[\left\|\mathbf{v}_k^{(q)}-\frac{Q}{Q-q}\,\mathrm{Log}_{\boldsymbol{\theta}_k^{(q)}}\!\left(\boldsymbol{\theta}_k^{\star}\right)\right\|_2^{2}\right]\\
&\quad+\beta_e\,\mathbb{E}\!\left[\left\|\mathrm{Log}_{\boldsymbol{\theta}_k^{(Q)}}\!\left(\boldsymbol{\theta}_k^{\star}\right)\right\|_2^{2}\right],
\end{aligned}
\end{equation}
where the second term penalizes the endpoint error with the weight $\beta_e$.
A small number of steps suffice to reach the refined latent $\boldsymbol{\theta}_k^{(Q)}$, which keeps the inference latency low.

The refined latent $\mathbf{z}_k^{(Q)}=\mathcal{M}\{\cos\theta_{k,j}^{(Q)},\sin\theta_{k,j}^{(Q)}\}_{j=1}^{J}$ and the reliability features $\mathbf{r}_k$ are passed to the semantic decoder $D_{\Omega}$ with parameters $\Omega$, which reconstructs the message through two branches.
The image branch upsamples the latent through cascaded ConvNeXt blocks and an output head to restore the scene image, while the state branch aligns the meta slots in a slot refiner and reads the target range, velocity, and angle through an analytic meta head together with the target class through a classification head.
The reconstructed semantic information is given by
\begin{equation}\label{eq:decode}
\hat{\mathbf{m}}_k=D_{\Omega}\!\left(\mathbf{z}_k^{(Q)},\mathbf{r}_k\right)=[\hat{\mathbf{i}}_k,\hat{\boldsymbol{\varsigma}}_k],
\end{equation}
where $\hat{\mathbf{i}}_k$ denotes the recovered image and $\hat{\boldsymbol{\varsigma}}_k$ collects the recovered range, velocity, angle, and class.


\subsection{End-to-End Training}\label{subsec:training}
\vspace{-0.2em}

\begin{algorithm}[!t]
\caption{End-to-End Training of CoSMIC}
\label{alg:cosmic_train}
\vspace{-0.1em}
\begin{algorithmic}[1]
\Require frame samples $\{(\mathbf{o}_{k-1},\mathbf{m}_k)\}$, covert budget $\epsilon$, reference pairs $\{\mathbf{q}_{k,j}\}_{j=1}^{J}$, steps $Q$, weights $\lambda_f$, $\beta_r$
\Ensure encoder $E_{\Theta}$, velocity field $\mathbf{v}_{\Psi}$, decoder $D_{\Omega}$
\State initialize $\Theta$, $\Psi$, $\Omega$
\While{not converged}
    \State form $\mathbf{m}_k=[\mathbf{i}_k,\boldsymbol{\varsigma}_k]$ by \eqref{eq:semantic_payload}
    \State produce $(\bar{c}_{k,j},\bar{s}_{k,j})$ and $\tilde{\tau}_{k,j}$ by \eqref{eq:convnext}--\eqref{eq:tx_heads}
    \State compute $\lambda_k$ and $(c_{k,j},s_{k,j})$ by \eqref{eq:budget_function}--\eqref{eq:projected_pair}
    \State form $\mathbf{b}_k^{(1)}$ by~\eqref{eq:frontend_fields} and synthesize $\mathbf{x}_k^{(1)}$ by~\eqref{eq:phase_mapping} and \eqref{eq:transmit_signal}
    \State obtain $\mathbf{y}_k^{b}$ by \eqref{eq:bob_receive}
    \State recover $\mathbf{z}_k^{(0)}$ and $\mathbf{r}_k$ by~\eqref{eq:mmse}--\eqref{eq:coarse_latent}
    \State set $\boldsymbol{\theta}_k^{\star}$ from $(\bar{c}_{k,j},\bar{s}_{k,j})$ and sample $t\sim\mathcal{U}(0,1)$
    \State form the bridge $\boldsymbol{\theta}_{k,t}$ by \eqref{eq:bridge} and evaluate $\mathcal{L}_f$ by \eqref{eq:reflow_risk}
    \State unroll $Q$ steps by \eqref{eq:reflow_step} to reach $\boldsymbol{\theta}_k^{(Q)}$ and evaluate $\mathcal{L}_r$ by \eqref{eq:reflow_roll}
    \State decode $\hat{\mathbf{m}}_k$ by \eqref{eq:decode}
    \State evaluate $\mathcal{L}_{\mathrm{sem}}$ by \eqref{eq:loss_sem} and $\mathcal{L}_{\mathrm{tot}}$ by \eqref{eq:loss_tot}
    \State update $(\Theta,\Psi,\Omega)$ by a gradient step on $\mathcal{L}_{\mathrm{tot}}$
\EndWhile
\State \Return $E_{\Theta}$, $\mathbf{v}_{\Psi}$, $D_{\Omega}$
\end{algorithmic}
\vspace{-0.3em}
\end{algorithm}

The encoder $E_{\Theta}$, the velocity field $\mathbf{v}_{\Psi}$, and the decoder $D_{\Omega}$ are optimized jointly in an end-to-end training procedure.
The training objective combines a semantic reconstruction term and a flow matching term.
The reconstruction term measures the distortion between the recovered semantic message and the source as
\begin{equation}\label{eq:loss_sem}
\mathcal{L}_{\mathrm{sem}}=
\mathbb{E}\!\left[
\mathcal{L}_{\mathrm{i}}(\hat{\mathbf{i}}_k,\mathbf{i}_k)
+\mathcal{L}_{\boldsymbol{\varsigma}}(\hat{\boldsymbol{\varsigma}}_k,\boldsymbol{\varsigma}_k)
\right].
\end{equation}
The image and state distortions are defined as
\begin{equation}\label{eq:image_loss_components}
\mathcal{L}_{\mathrm{i}}(\hat{\mathbf{i}}_k,\mathbf{i}_k)
=\frac{1}{N_{\mathrm{i}}}\|\hat{\mathbf{i}}_k-\mathbf{i}_k\|_2^2,
\end{equation}
and
\begin{equation}\label{eq:state_loss_components}
\begin{aligned}
&\mathcal{L}_{\boldsymbol{\varsigma}}(\hat{\boldsymbol{\varsigma}}_k,\boldsymbol{\varsigma}_k)
=\frac{1}{L_k}\sum_{\ell=1}^{L_k}
\Biggl[
\left(\frac{\hat{r}_{k,\ell}-r_{k,\ell}}{R_{\max}}\right)^2\\
&+\left(\frac{\hat{v}_{k,\ell}-v_{k,\ell}}{V_{\max}}\right)^2
+\left(\frac{\hat{\theta}_{k,\ell}-\theta_{k,\ell}}{\Theta_{\max}}\right)^2
-\log \hat{\pi}_{k,\ell,g_{k,\ell}}
\Biggr],
\end{aligned}
\end{equation}
where $N_{\mathrm{i}}$ denotes the number of image pixels, $R_{\max}$, $V_{\max}$, and $\Theta_{\max}$ denote the normalization ranges, and $\hat{\pi}_{k,\ell,g_{k,\ell}}$ denotes the predicted probability of class $g_{k,\ell}$.
The flow matching term combines the velocity risk in~\eqref{eq:reflow_risk} and the rollout consistency risk in~\eqref{eq:reflow_roll} evaluated by Monte Carlo sampling of the bridge time, and the total loss is
\begin{equation}\label{eq:loss_tot}
\mathcal{L}_{\mathrm{tot}}=\mathcal{L}_{\mathrm{sem}}+\lambda_f{\left(\mathcal{L}_f+\beta_r\mathcal{L}_r\right)},
\end{equation}
where $\lambda_f$ weights the flow matching term and $\beta_r$ weights the rollout consistency risk.
The complete training procedure is summarized in Algorithm~\ref{alg:cosmic_train}.



\vspace{-0.5em}

\section{Numerical and Simulation Results}\label{sec:results}
\vspace{-0.5em}

In this section, numerical and simulation results are presented to evaluate the performance of the proposed system.

\subsection{Simulation Settings}\label{subsec:sim_settings}

\subsubsection{Dataset}
To align with the vehicular ISAC scenarios, we adopt the View-of-Delft (VoD) automotive dataset~\cite{palffy2022vod} with $8693$ urban traffic frames that contain synchronized camera images, LiDAR labels and $77$~GHz radar measurements.
Each frame supplies the scene image together with the target range, velocity, angle, and class.
The official training and validation splits of VoD are built from separate driving sequences, and the labels of the official test split are withheld.
The networks are trained on $4000$ Car targets of the training split selected by bounding-box size and visibility.
Each image is the target region cropped from the $1936\times1216$ camera frame and fitted to a $192\times320$ canvas with the aspect ratio kept.
Each point of the performance curves averages at least $480$ test samples under channel realizations drawn with a fixed seed.

\subsubsection{Network Settings}
The semantic encoder stacks a stem layer and $6$ ConvNeXt blocks, while the decoder employs $6$ latent blocks and $4$ image blocks.
The velocity field of the RFlow refiner is a UNet with $12.4$~M parameters and runs $Q=2$ integration steps with the Euler solver.
The detailed configuration is listed in Table~\ref{tab:network_config}.

\subsubsection{Training Parameters}
All models are trained for $30$ epochs by the AdamW optimizer with a learning rate of $3\times10^{-4}$.
The SemCom channel follows Rayleigh fading, and the training SNR is drawn continuously over $0$ to $20$~dB.
All models are trained and timed on an NVIDIA RTX 4090 D GPU under PyTorch 2.5.1, where the latencies are measured at a batch size of one by CUDA event timing.

\subsubsection{Benchmarks}
Three representative schemes are adopted for comparison.
Specifically, HPQ~\cite{yilmaz2024high} refines coarse images via receiver-side diffusion, CDDM~\cite{wu2024cddm} integrates post-equalization channel denoising, and SIMAC~\cite{peng2026simac} implements ISAC through multimodal semantic fusion.
HPQ and CDDM share the transmitter and the receiver front end of CoSMIC and replace the RFlow refiner with their own diffusion modules, where HPQ restores the decoded image through the null-space restoration and CDDM denoises the equalized symbols at the timestep matched to the channel SNR.
SIMAC keeps its multimodal encoder and decoder and embeds its channel symbols as a residual on a public cover waveform under the same KL budget, where its state branch carries the eight state entries in $256$ dedicated complex symbols at a four-fold symbol power.
The budgets are matched in the real channel dimensions per frame under the same transmit power, i.e., $15360$ phases of the constant-modulus samples for CoSMIC and $7936$ complex symbols for SIMAC.
All learnable modules of the benchmarks are retrained jointly under the channel model, SNR range, and training schedule of CoSMIC.


\subsubsection{System and Radar Parameters}
The dual-functional waveform operates at a carrier frequency of $f_c=77$~GHz with a sweep bandwidth of $B=150$~MHz.
Each frame consists of $M=64$ chirps of $N=240$ samples with a chirp repetition interval of $T_r=40$~$\mu$s and therefore occupies $MN=15360$ channel uses, which carry $J=7680$ Gaussian pairs at a bandwidth ratio of $1/12$ with respect to the $192\times320\times3$ scene image.
Alice collects echoes with a uniform linear array of $U=8$ elements at half-wavelength spacing over an angular field of view of $[-60^\circ,60^\circ]$.
The phase mapping gain is set to $\alpha_{\phi}=0.26$, $\lambda_f$ in~\eqref{eq:loss_tot} is set to $0.36$, and the covertness constraint is fixed at $\epsilon=0.10$ unless stated otherwise.
The reserved groups contain $192$, $224$, and $640$ pairs for the range, velocity, and angle, $288$ pairs for the bounding box of the target in the image, and $96$ pairs for the class, where the state slots of the first four components amount to $1344$ pairs, i.e., $17.5\%$ of the $J=7680$ pairs.

\begin{table}[t]
\centering
\caption{Layer Configuration of the Proposed Network}
\label{tab:network_config}
\vspace{-0.4em}
\renewcommand{\arraystretch}{1.0}
\setlength{\tabcolsep}{4pt}
\begin{tabular}{llcc}
\hline
\textbf{Stage} & \textbf{Layer} & \textbf{Width} & \textbf{Act.} \\
\hline
\multirow{4}{*}{Encoder}
 & Stem (stride-$4$ conv)     & $80$   & GELU \\
 & ConvNeXt block $\times 6$   & $80$   & GELU \\
 & Meta slot injection        & --     & --   \\
 & Direction / scale head     & $2/1$  & $\ell_2$ / Softplus \\
\hline
\multirow{2}{*}{RFlow Refiner}
 & UNet with attention & $128$--$256$ & SiLU \\
 & Velocity head              & $1$    & --   \\
\hline
\multirow{5}{*}{Decoder}
 & ConvNeXt block $\times 6$   & $224$  & GELU \\
 & Slot refiner               & $192$  & GELU \\
 & Meta / class head          & $4/13$ & -- / Softmax \\
 & ConvNeXt block $\times 4$   & $224$  & GELU \\
 & Upsample ($4\times$) $+$ head & $3$ & Sigmoid \\
\hline
\end{tabular}
\vspace{-0.4em}
\end{table}

\subsection{Covertness and Sensing Performance}
\vspace{-0.25em}

\begin{figure*}[!t]
\centering
\subfigure[]{\includegraphics[width=0.245\linewidth]{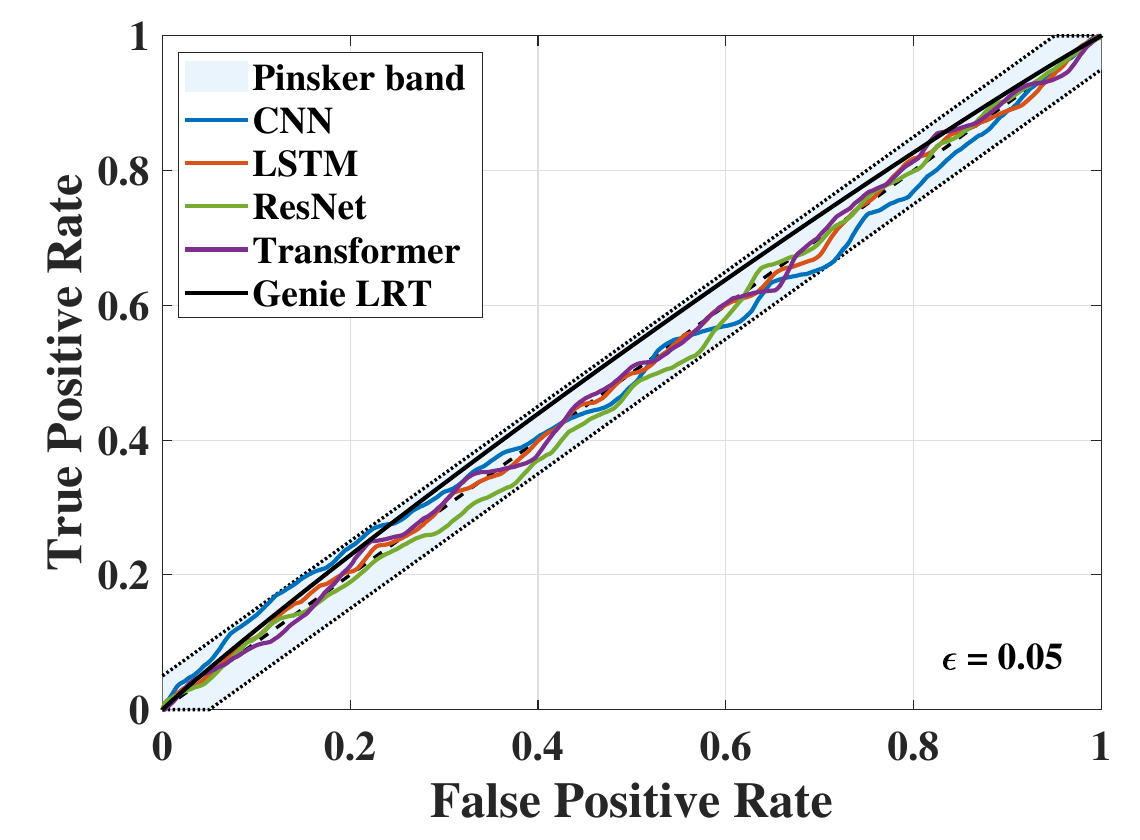}\label{willie_roc_eps005}}\hfill
\subfigure[]{\includegraphics[width=0.245\linewidth]{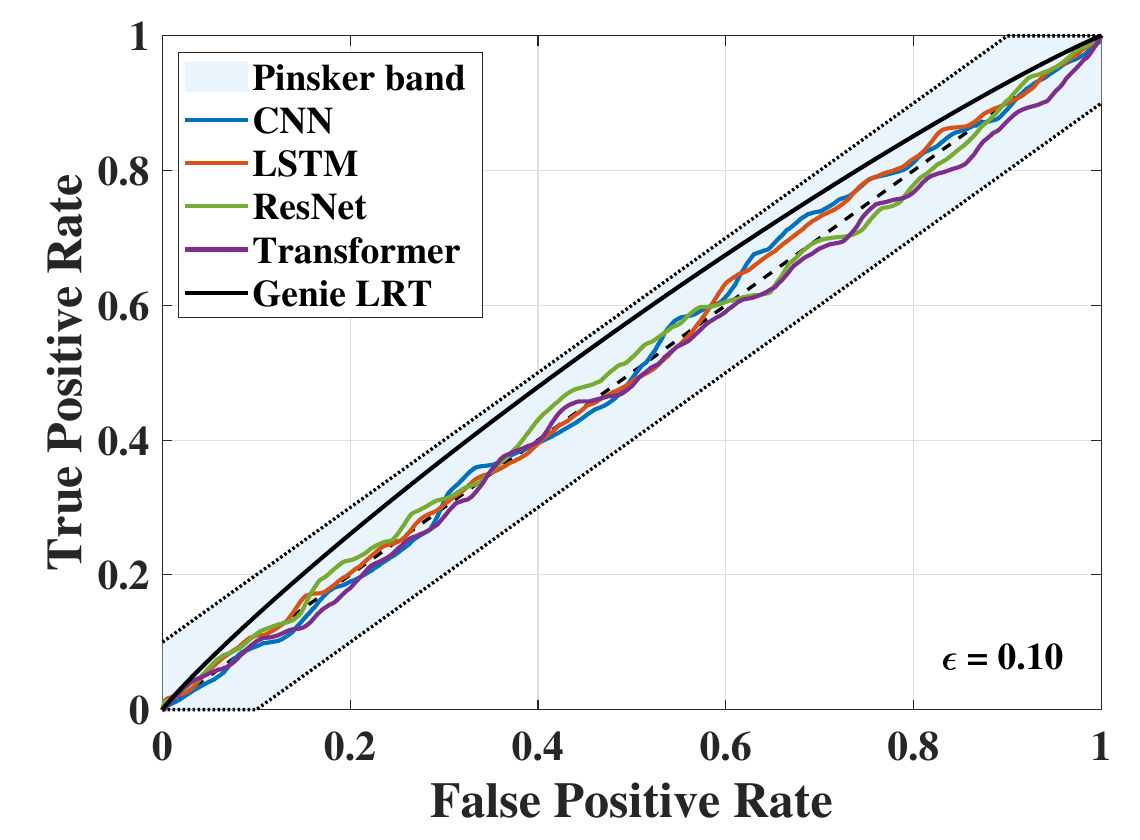}\label{willie_roc_eps010}}\hfill
\subfigure[]{\includegraphics[width=0.245\linewidth]{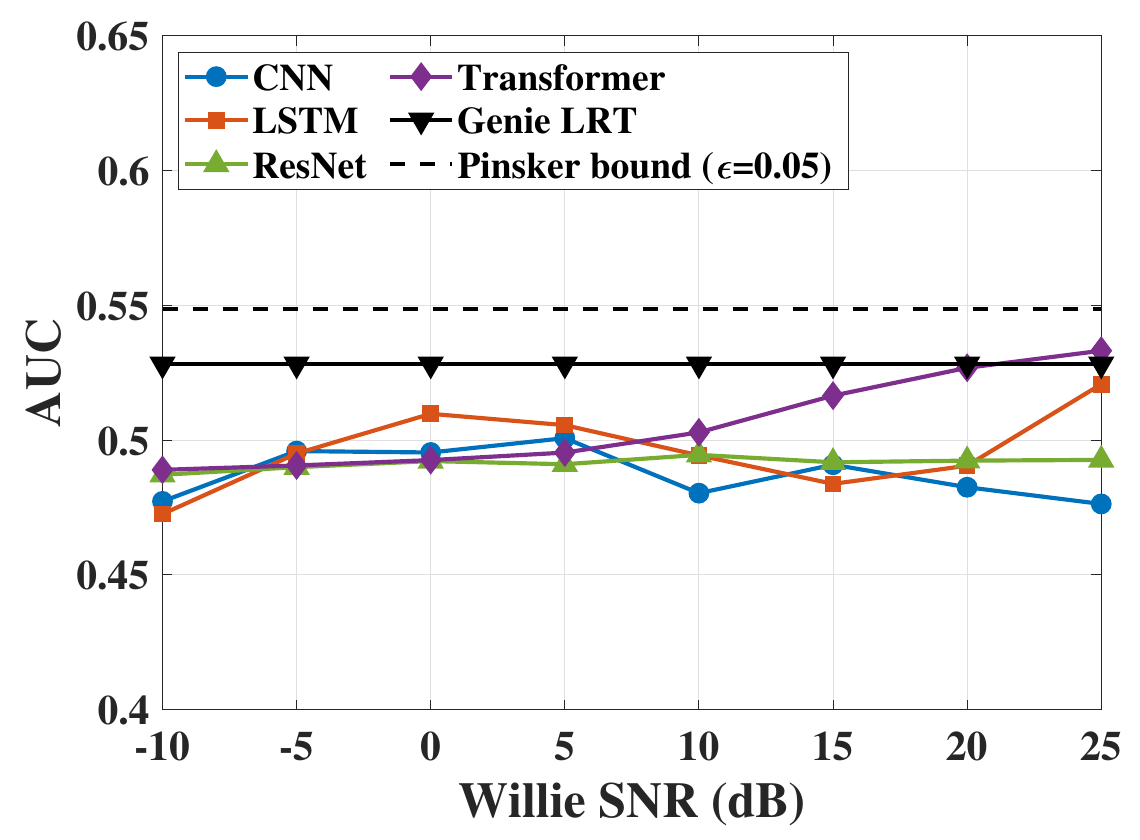}\label{willie_auc_eps005}}\hfill
\subfigure[]{\includegraphics[width=0.245\linewidth]{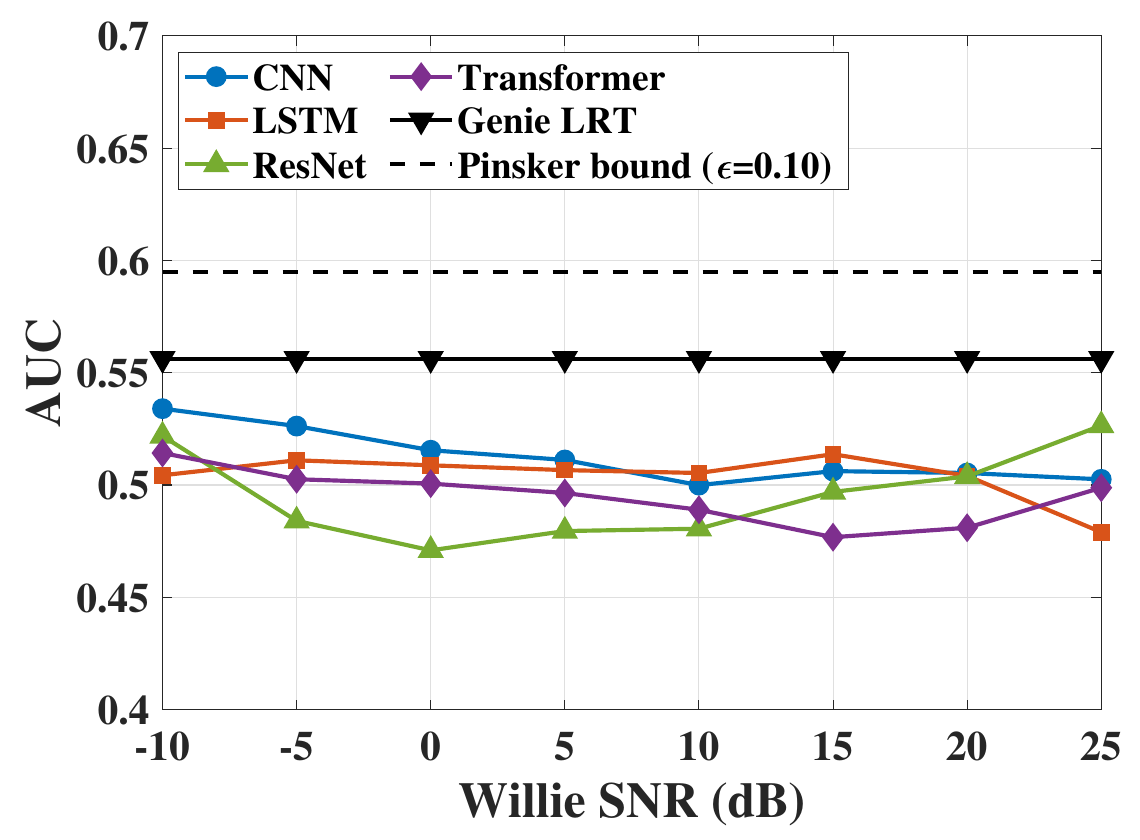}\label{willie_auc_eps010}}%
\vspace{-0.5em}
\caption{Detection performance of Willie over a monitoring window of $K_{\mathrm{W}}=1000$ frames under the learning-based detectors in~\cite{qian2025adversarial} and the genie LRT. (a) and (b) depict the ROC at a Willie SNR of $10$~dB, while (c) and (d) depict the AUC versus the Willie SNR. The shaded region and the dashed line denote the Pinsker bound associated with the covertness budget $\epsilon$.}
\label{willie_metrics}
\vspace{-0.8em}
\end{figure*}

\begin{figure*}[!t]
\centering
\subfigure[]{\includegraphics[width=0.328\linewidth]{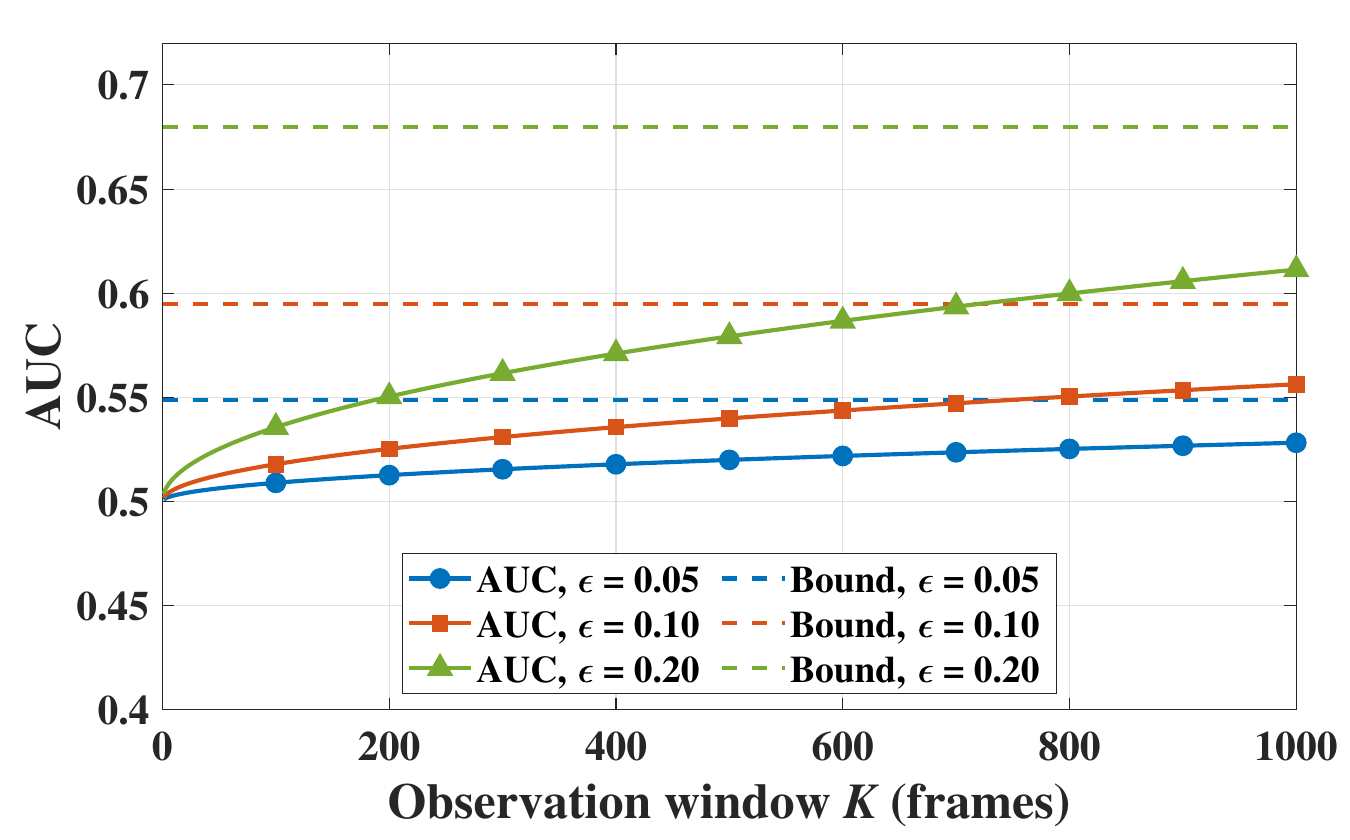}\label{auc_window}}\hfill
\subfigure[]{\includegraphics[width=0.328\linewidth]{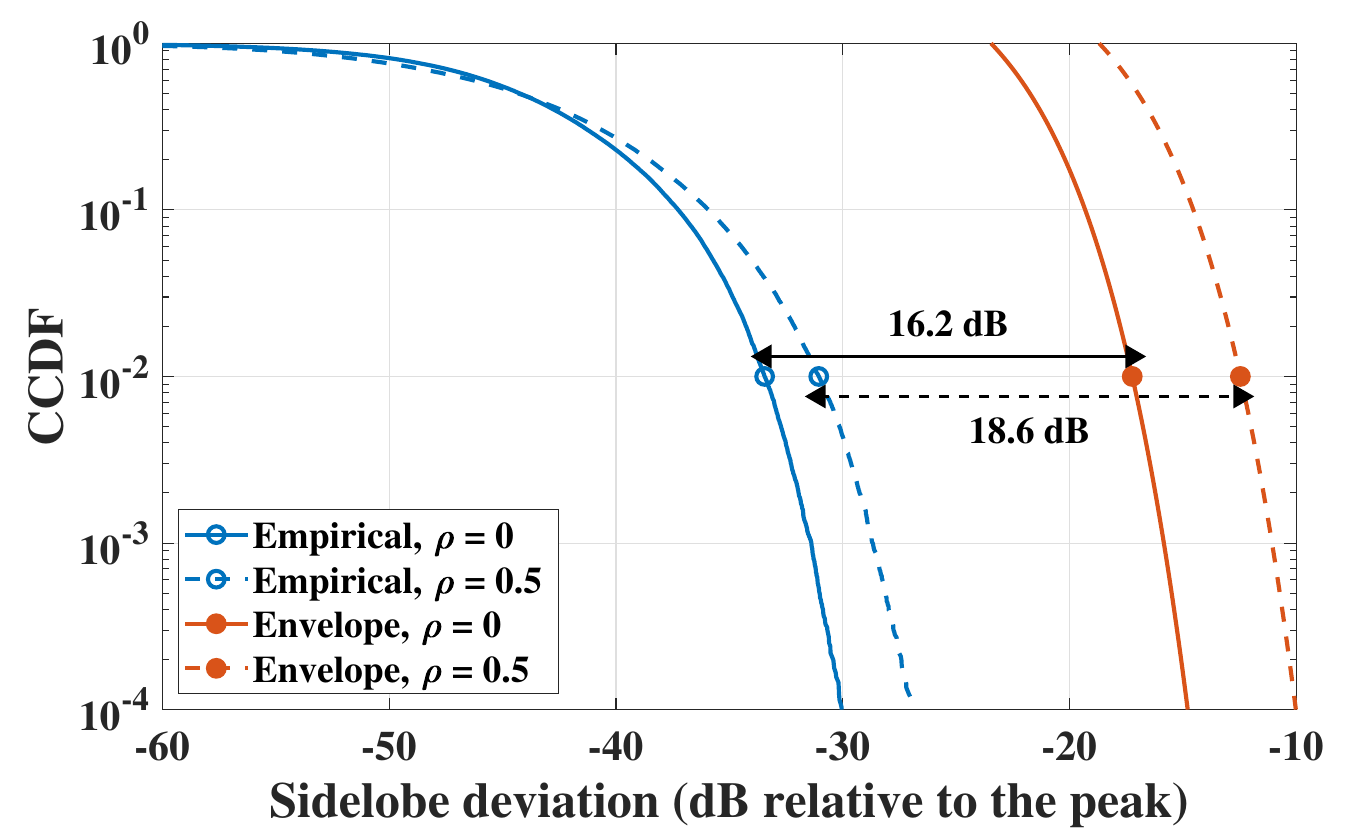}\label{envelope_ccdf}}\hfill
\subfigure[]{\includegraphics[width=0.328\linewidth]{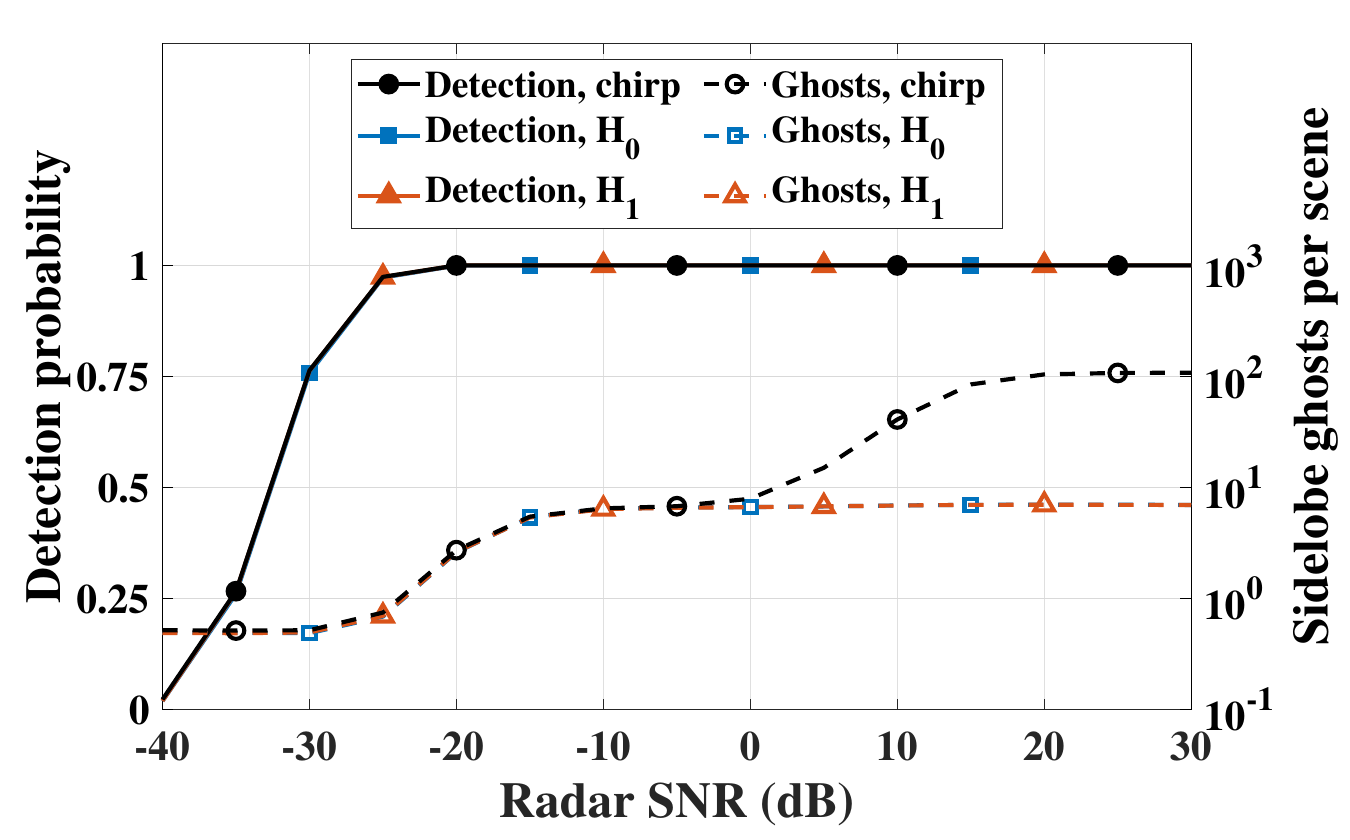}\label{multitarget_detection}}%
\vspace{-0.9em}
\caption{Multi-frame covertness and sensing validation. (a) AUC of the genie LRT versus the observation window $K$ with $K_{\mathrm{W}}=1000$, where the dashed lines denote the Pinsker bounds. (b) Empirical CCDF of the inter-mode sidelobe deviation against the certified envelope $B_{E,k}$, where the markers denote $\eta=0.01$. (c) Detection probability~(left axis) and sidelobe ghosts per scene~(right axis) of the unmodulated chirp and the $H_0$ and $H_1$ waveforms under the same power and bandwidth.}
\label{covert_sensing}
\vspace{-0.8em}
\end{figure*}

To assess the covertness under intelligent detection, Willie is equipped with the DL-based detectors in~\cite{qian2025adversarial}, which adopt the convolutional neural network~(CNN), long short-term memory~(LSTM), residual network~(ResNet), and Transformer architectures to classify the observations over a monitoring window of $K_{\mathrm{W}}=1000$ consecutive frames into $H_0$ and $H_1$.
Each detector takes the in-phase and quadrature~(I/Q) samples of a received frame as the input, where the CNN stacks four convolutional blocks, the ResNet stacks three residual blocks, and the LSTM and the Transformer process the chirps through two layers with a hidden dimension of $128$.
For every covertness budget and every Willie SNR, the detectors are trained on labeled frames of both transmit modes by the Adam optimizer with the binary cross-entropy loss, where the training and test frames are drawn from disjoint subsets of the dataset.
The frame-level decision statistics of each detector are accumulated over the window, and the closed-form genie likelihood ratio test~(LRT) with knowledge of the scale map of every frame is included as the reference detector.
Fig.~\ref{willie_roc_eps005} and~\ref{willie_roc_eps010} present the receiver operating characteristic~(ROC) of the detectors at a Willie SNR of $10$~dB for $\epsilon=0.05$ and $\epsilon=0.10$, respectively.
The ROC of an arbitrary detector remains inside the Pinsker band enclosing the random guessing diagonal, which indicates that the architectures with richer features obtain no appreciable detection advantage.
Fig.~\ref{willie_auc_eps005} and~\ref{willie_auc_eps010} further report the area under the curve~(AUC) of the detectors against the Willie SNR from $-10$ to $25$~dB.
For both covertness budgets, the AUC of each learning-based detector fluctuates around $0.5$ and stays below the Pinsker bound over the entire range, which indicates that a warden obtaining a cleaner observation gains no additional ability to separate $H_1$ from $H_0$.
Meanwhile, the AUC of the genie LRT stays constant over the Willie SNR and lies below the Pinsker bound.
The results confirm that the window allocation under Theorem~\ref{thm:covertness} confines the detection performance of Willie over the monitoring window within the prescribed covertness budget.
\looseness=-1

Fig.~\ref{auc_window} presents the AUC of the genie LRT versus the observation window $K$ under the window allocation with $K_{\mathrm{W}}=1000$, together with the Pinsker bounds $1/2+\epsilon-\epsilon^{2}/2$ of the covertness budgets.
The AUC increases mildly with the observation window and remains below the bound of the corresponding budget for every window length, which confirms that the window allocation keeps the joint observation of consecutive frames within the prescribed covertness budget.

\begin{figure}[t]
  \centering
  \includegraphics[width=0.95\linewidth]{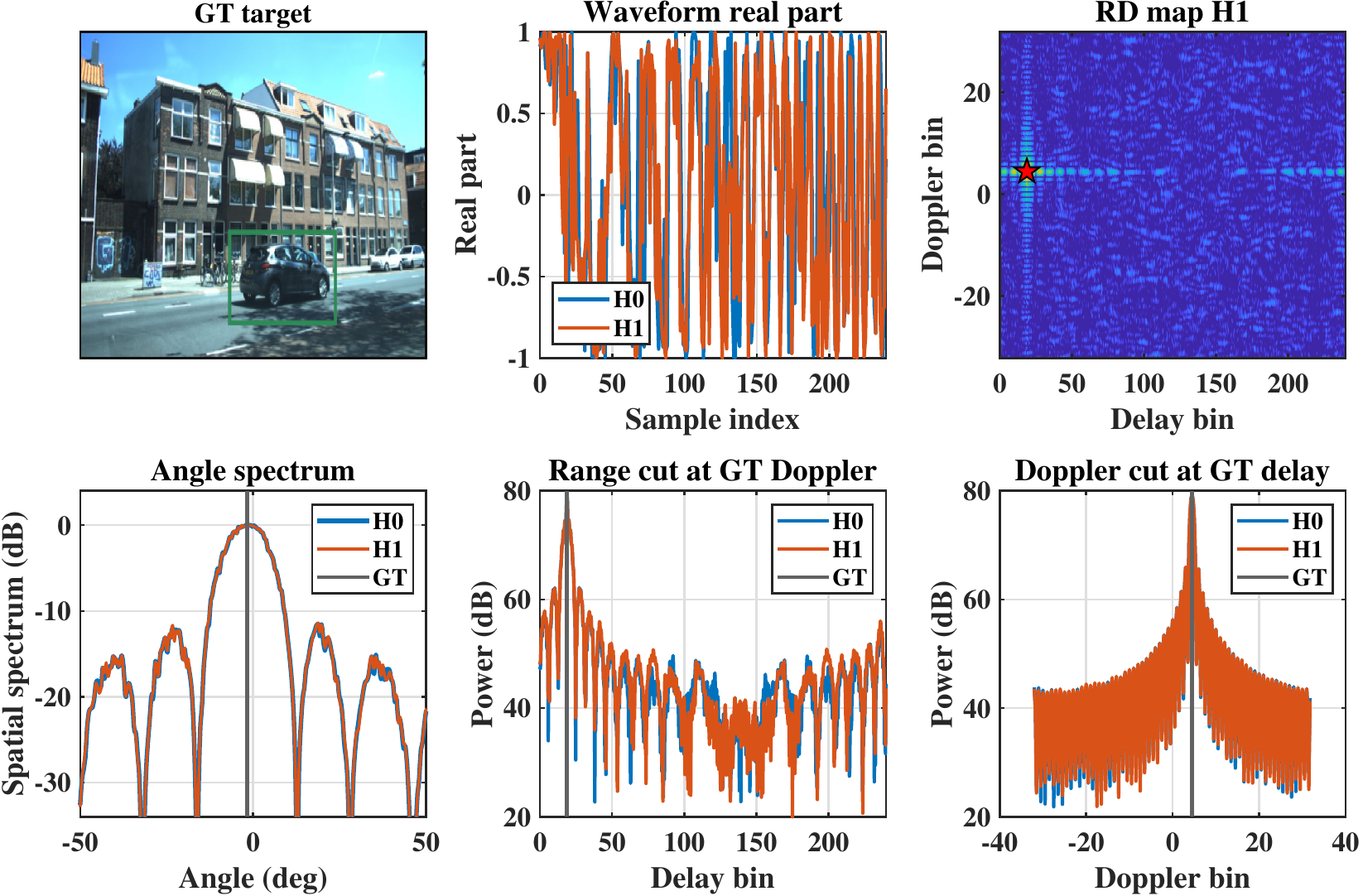}
\vspace{-0.4em}
  \caption{Radar responses under the sensing-only waveform $H_0$ and the covert semantic waveform $H_1$ for a representative frame.}
  \label{radar_h0_h1_insights}
\vspace{-0.6em}
\end{figure}

\begin{figure}[t]
  \centering
  \includegraphics[width=0.8\linewidth]{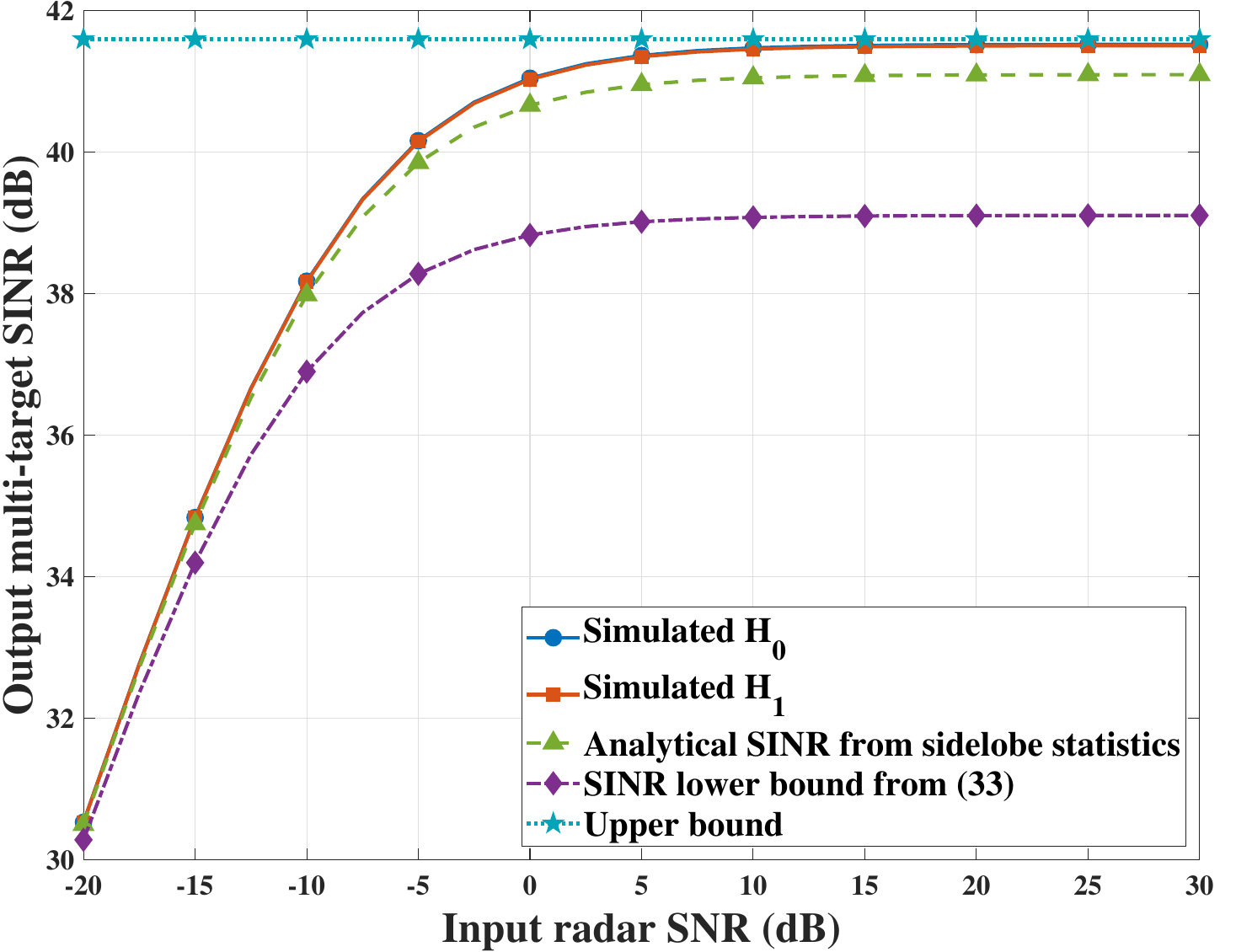}
\vspace{-0.4em}
  \caption{Comparison of the output multi-target SINR of radar versus input radar SNR under hypotheses $H_0$ and $H_1$ with the analytical SINR and the lower bound in~\eqref{eq:sinr_bound}.}
  \label{multitarget_sinr_vs_snr}
\vspace{-0.6em}
\end{figure}

To intuitively illustrate the covertness at the waveform level, Fig.~\ref{radar_h0_h1_insights} compares the transmit waveform and the matched responses of the sensing-only mode $H_0$ and the covert mode $H_1$ for a representative frame.
The real part of the transmit waveform under $H_0$ and $H_1$ overlaps throughout the sample index, and the constant-modulus envelope is retained.
The range-Doppler map under $H_1$ resolves the target at the true bin, while the range, Doppler, and angle responses under $H_0$ and $H_1$ coincide and align with the ground truth~(GT), which indicates that the embedding of the semantic message leaves the second-order statistics and the spectral occupancy of the waveform virtually unchanged.
In the frequency domain, the power spectral densities of the waveforms under $H_0$ and $H_1$ differ by at most $0.14$~dB over the sweep bandwidth, and the $99\%$ occupied bandwidth equals $148$~MHz under both modes.
Since the discriminative features extracted from the waveform remain effectively indistinguishable, a multi-feature DL-based detector at Willie obtains no separable evidence between $H_0$ and $H_1$.
Therefore, the sensing accuracy and the covertness are attained simultaneously.


%
\begin{figure*}[!t]
  \centering
  \subfigure[PSNR of the reconstructed image versus the SNR of Bob.]{
      \includegraphics[width=0.31\textwidth]{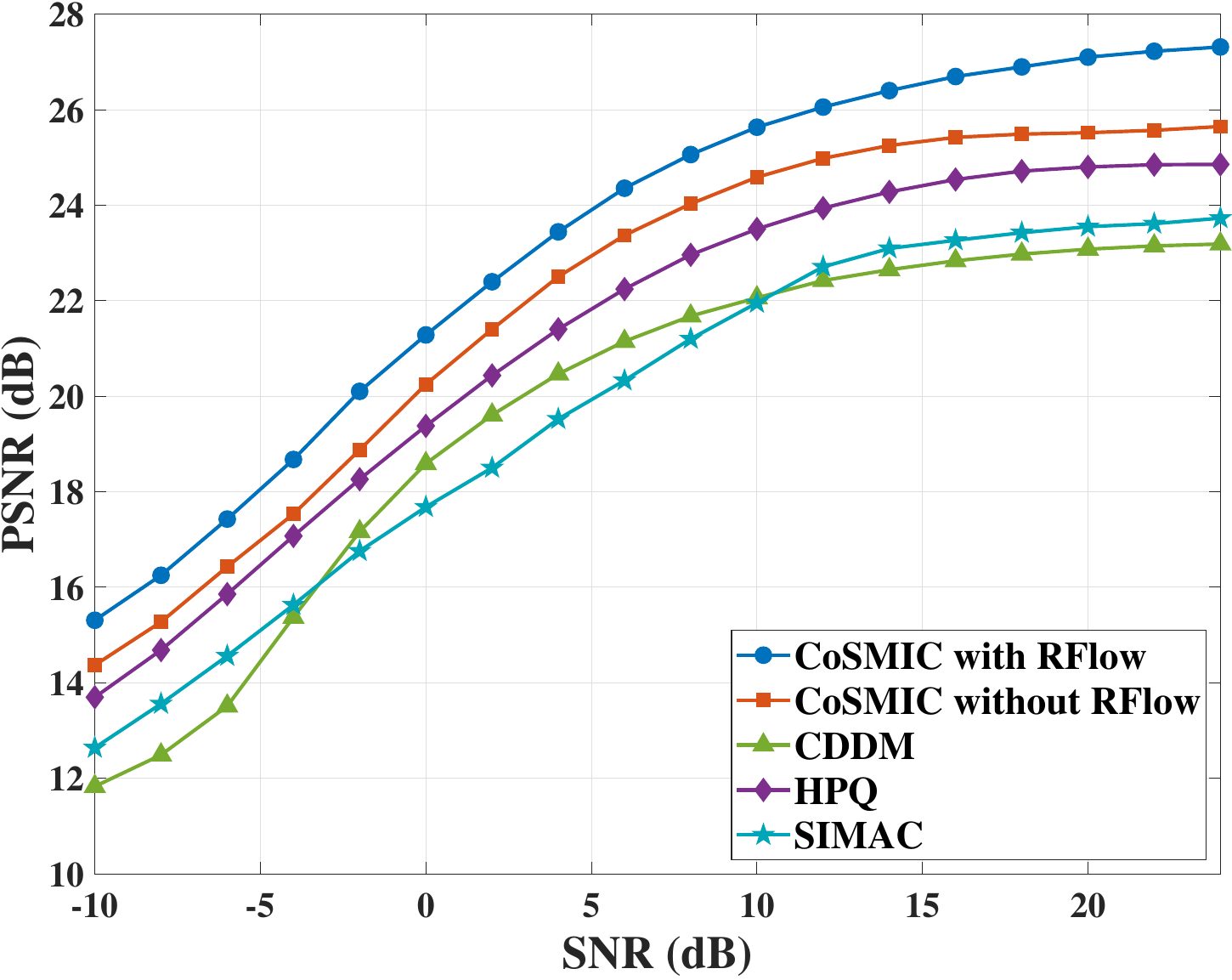}
      \label{psnr_comparison}
  }
  \hfil
  \subfigure[SSIM of the reconstructed image versus the SNR of Bob.]{
      \includegraphics[width=0.312\textwidth]{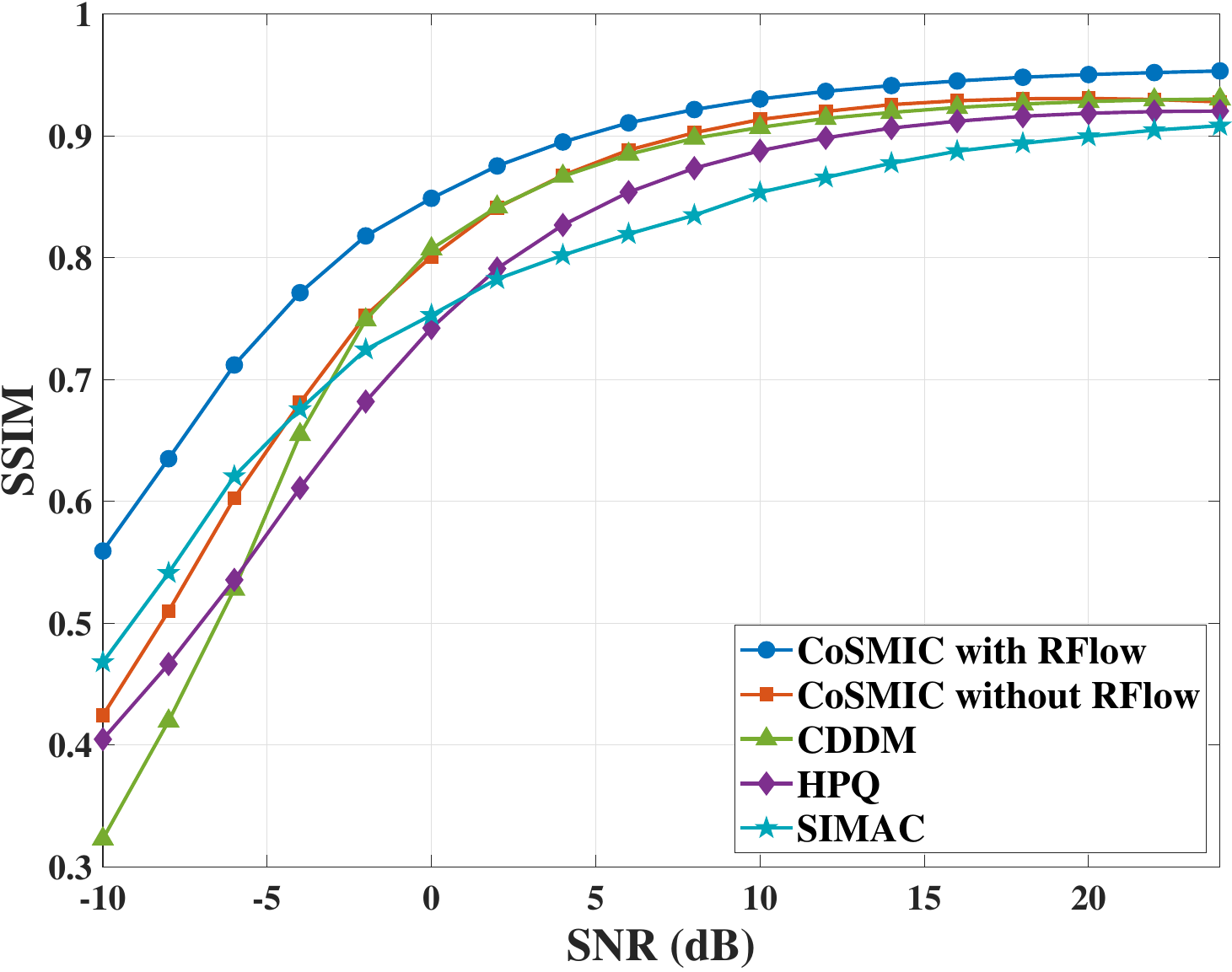}
      \label{ssim_comparison}
  }
  \hfil
  \subfigure[RMSE of the recovered range, velocity, and angle versus the SNR of Bob.]{
      \includegraphics[width=0.308\textwidth]{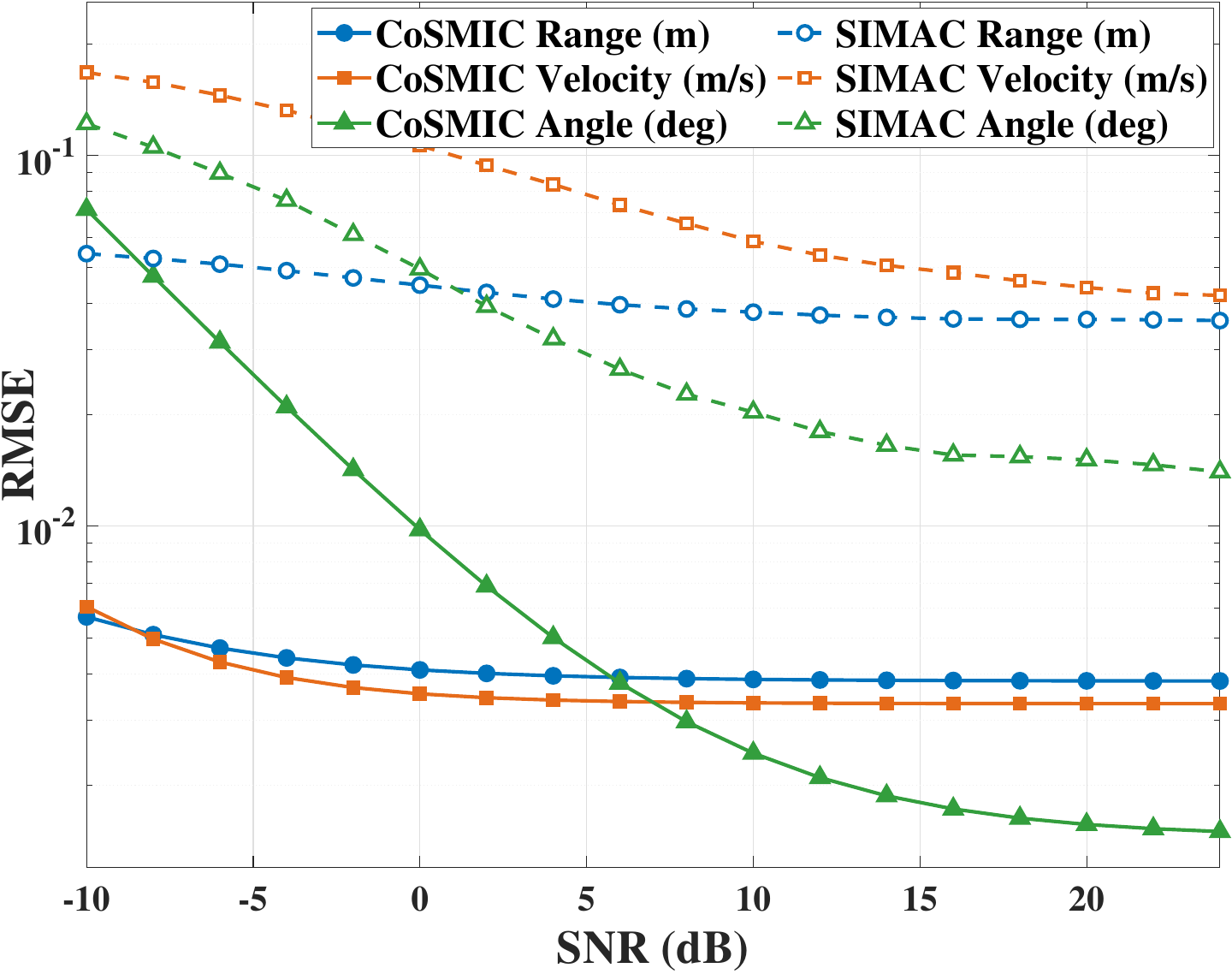}
      \label{radar_rmse}
  }
\vspace{-0.4em}
  \caption{SemCom performance comparison of CoSMIC and the benchmark schemes CDDM~\cite{wu2024cddm}, HPQ~\cite{yilmaz2024high}, and SIMAC~\cite{peng2026simac} over a Rayleigh fading channel.}
  \label{performance_comparison}
\vspace{-0.8em}
\end{figure*}

\begin{figure*}[!t]
    \centering
    \includegraphics[width=0.86\textwidth]{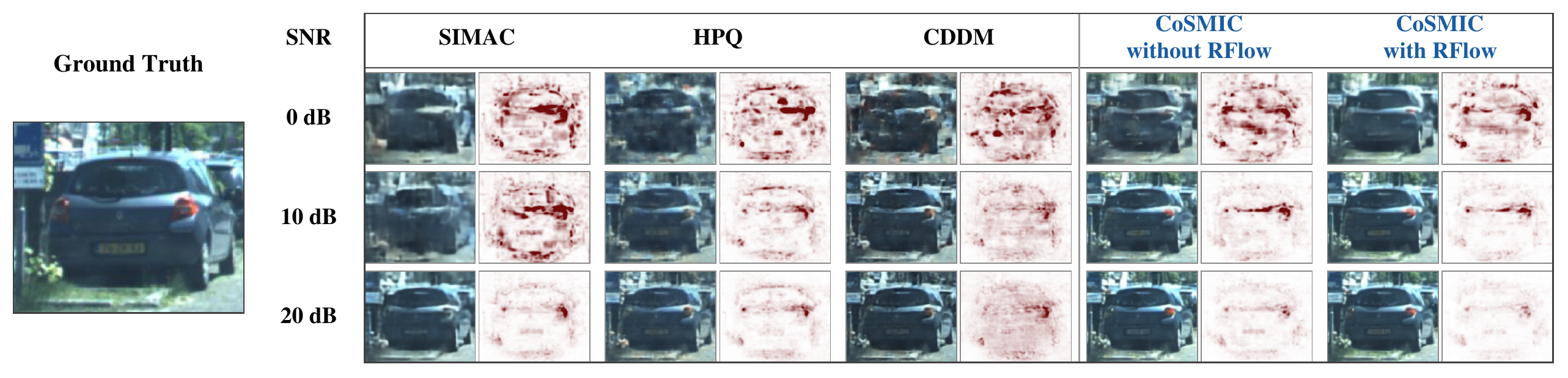}
\vspace{-0.5em}
    \caption{Reconstructed images and residual maps of CoSMIC and the benchmark schemes at various SNRs.
}
    \label{semantic_examples}
\vspace{-0.8em}
\end{figure*}

Fig.~\ref{multitarget_sinr_vs_snr} plots the output multi-target SINR against the input radar SNR for the simulated $H_0$ and $H_1$, the analytical SINR, the lower bound in~\eqref{eq:sinr_bound}, and the noise-free upper bound.
The output SINR increases steadily with the input SNR in the noise-limited regime and saturates toward the interference-limited ceiling represented by the noise-free upper bound at a high input SNR.
The simulated SINR under $H_0$ and $H_1$ remain identical over the whole range, which indicates that the covert embedding leaves the sensing function intact.
Furthermore, the analytical SINR from the sidelobe statistics matches the simulated curves, while the lower bound in~\eqref{eq:sinr_bound} remains consistently below both curves, confirming the validity of the derived bound across the entire SNR range.

Fig.~\ref{envelope_ccdf} verifies the envelope $B_{E,k}$ through the empirical complementary cumulative distribution function~(CCDF) of the inter-mode sidelobe deviation $|A_k^{(1)}-A_k^{(0)}|$ over $2000$ frames and $48$ lags for $\rho=0$ and $\rho=0.5$.
The empirical tail lies inside the certified envelope at every confidence level with zero violations at $\eta\in\{0.01,0.05,0.1\}$, where the margin reflects the worst case over all rotation fields and scale maps.

Fig.~\ref{multitarget_detection} compares the multi-target sensing performance of the unmodulated chirp and the $H_0$ and $H_1$ waveforms under the same transmit power and sweep bandwidth, where each scene contains two, three, or five targets with random off-grid ranges, velocities, and angles.
The targets are detected by a cell-averaging constant false alarm rate~(CA-CFAR) detector with $P_{\mathrm{fa}}=10^{-4}$, and the sidelobe ghosts denote the false detections produced by the sidelobes of the detected targets.
The detection probability curves of the three waveforms coincide, where the SNR required for a detection probability of $0.9$ differs by less than $0.1$~dB.
From a radar SNR of $0$~dB, the deterministic range sidelobes of the chirp raise the ghost count of the chirp above a hundred per scene at $20$~dB, whereas the random phase coding of CoSMIC converts the range sidelobes into a noise-like floor and keeps the ghosts of $H_0$ and $H_1$ at a low common level.
Moreover, the range, velocity, and angle root mean square errors~(RMSEs) of the matched detections coincide for $H_0$ and $H_1$.
At an SNR of $10$~dB, two targets separated by two or three resolution cells are resolved with a probability above $0.96$, which confirms that the semantic embedding preserves the detection and estimation performance.
\looseness=-1

\subsection{SemCom Performance}
\vspace{-0.25em}

Fig.~\ref{psnr_comparison} shows the peak signal-to-noise ratio~(PSNR) of CoSMIC and the benchmark schemes versus the SNR from $-10$ to $24$~dB over the Rayleigh channel.
CoSMIC with RFlow refiner delivers the highest PSNR over the considered SNR range and surpasses the benchmark schemes by $10\%$ to $18\%$.
Moreover, the advantage is preserved as the SNR decreases, since the covert rotation coding places the semantics on the phase directions of a constant-modulus waveform, and the directional latent therefore stays informative under strong noise.
The reliability-guided refinement yields a further gain of $1$ to $2$~dB over CoSMIC without RFlow, as the velocity field steers the corrupted latent back toward the clean manifold under the guidance of the per-pair reliability.

\begin{figure*}[!t]
\centering
\subfigure[]{\includegraphics[width=0.2398\linewidth]{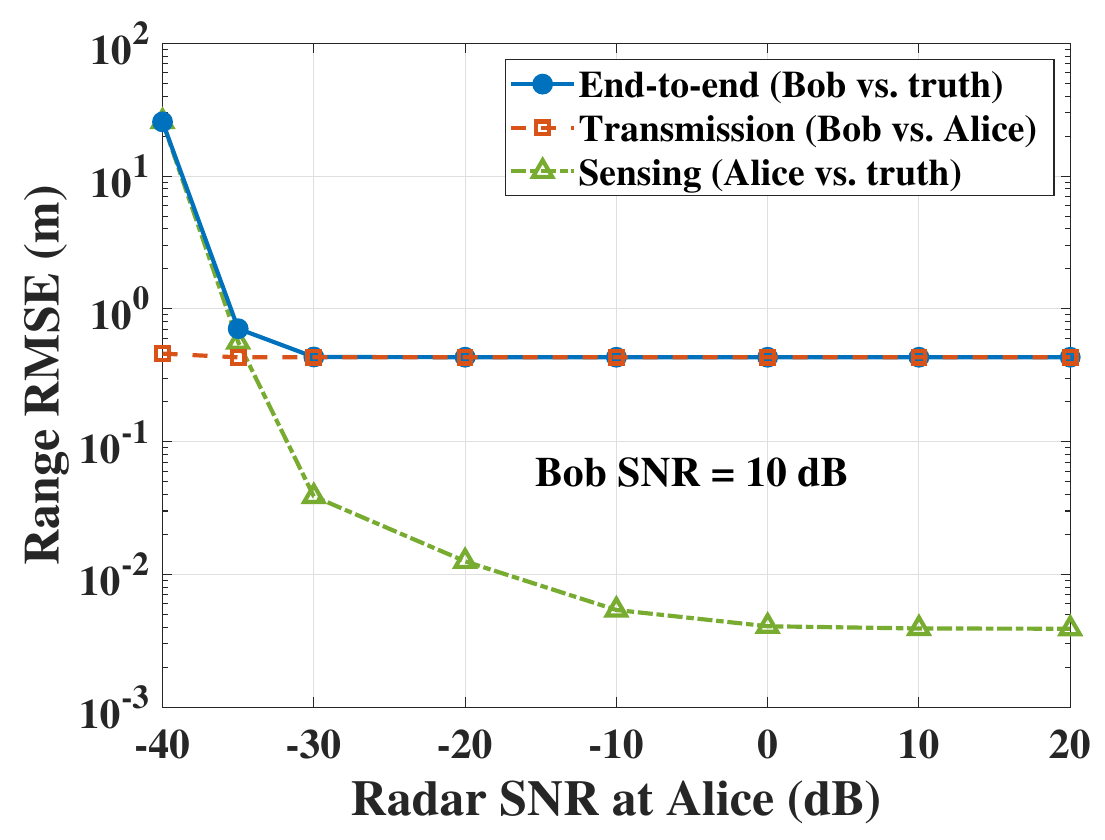}\label{state_error_decomposition}}\hfill
\subfigure[]{\includegraphics[width=0.2584\linewidth]{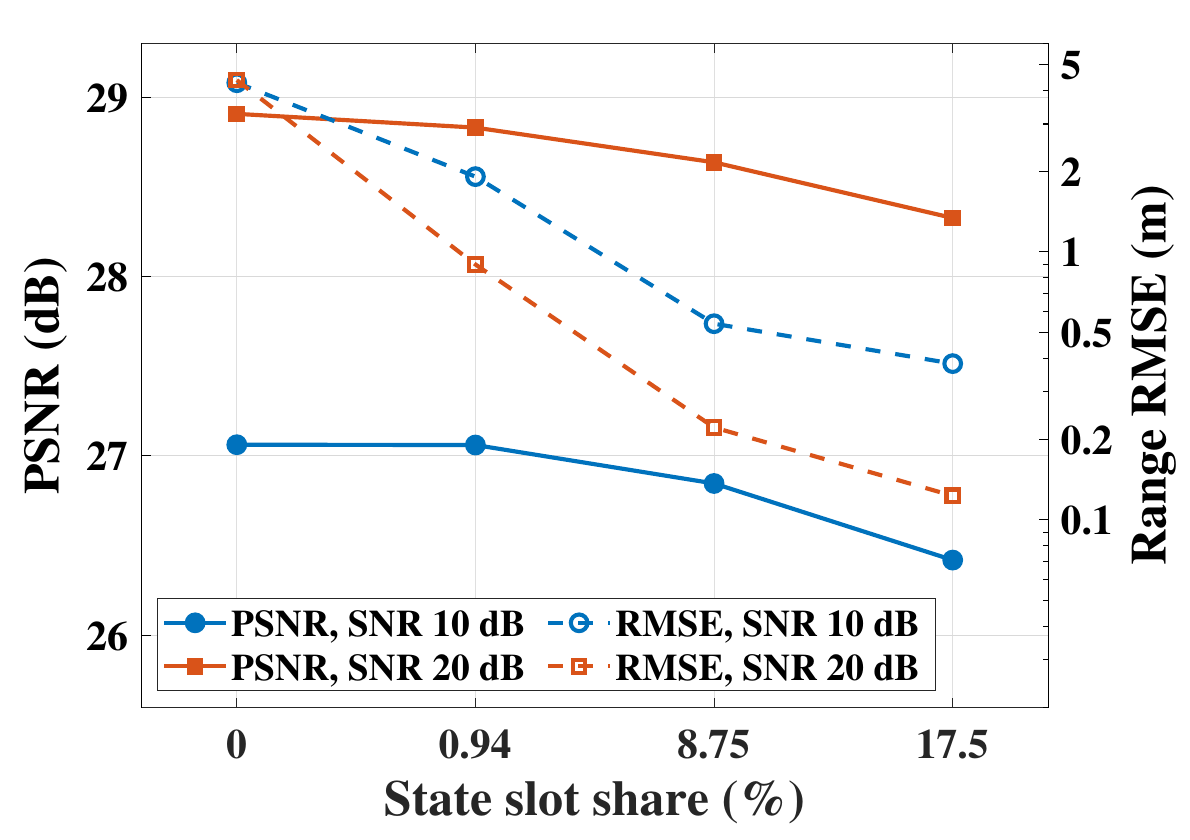}\label{slot_share_tradeoff}}\hfill
\subfigure[]{\includegraphics[width=0.2561\linewidth]{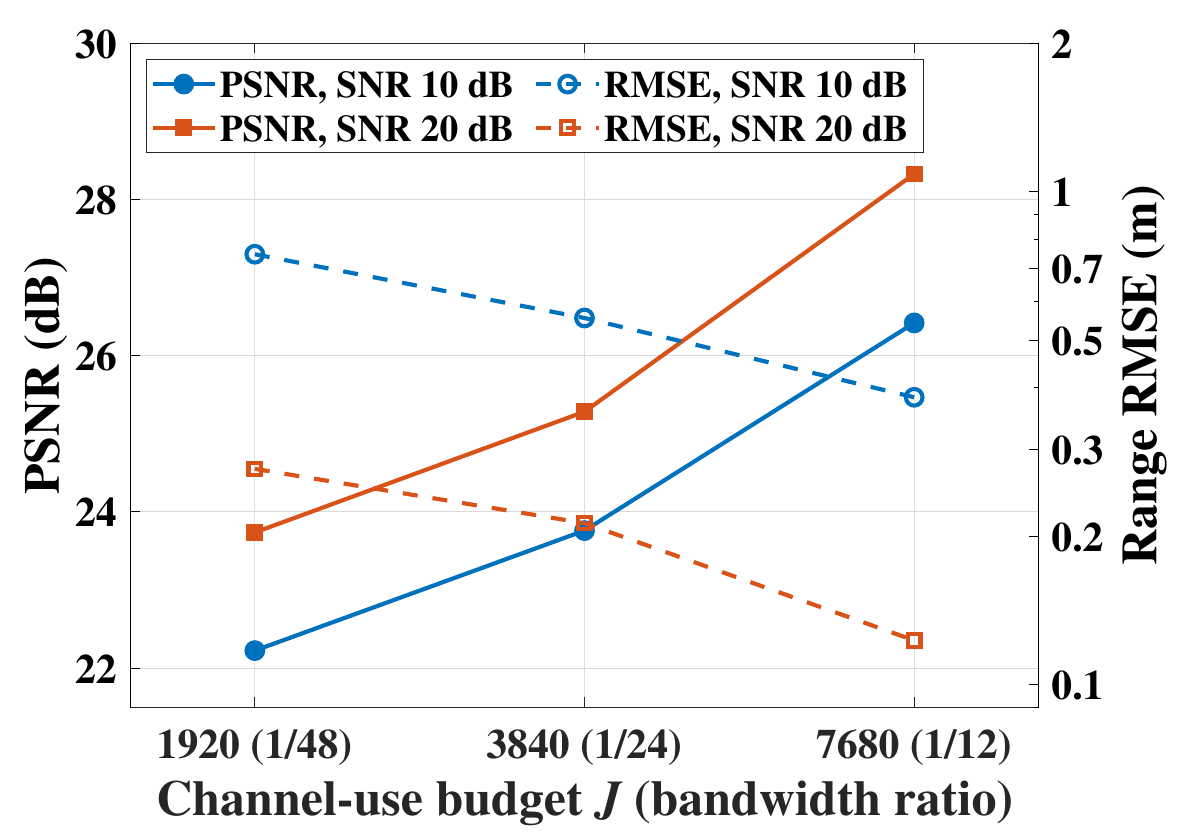}\label{budget_tradeoff}}\hfill
\subfigure[]{\includegraphics[width=0.2353\linewidth]{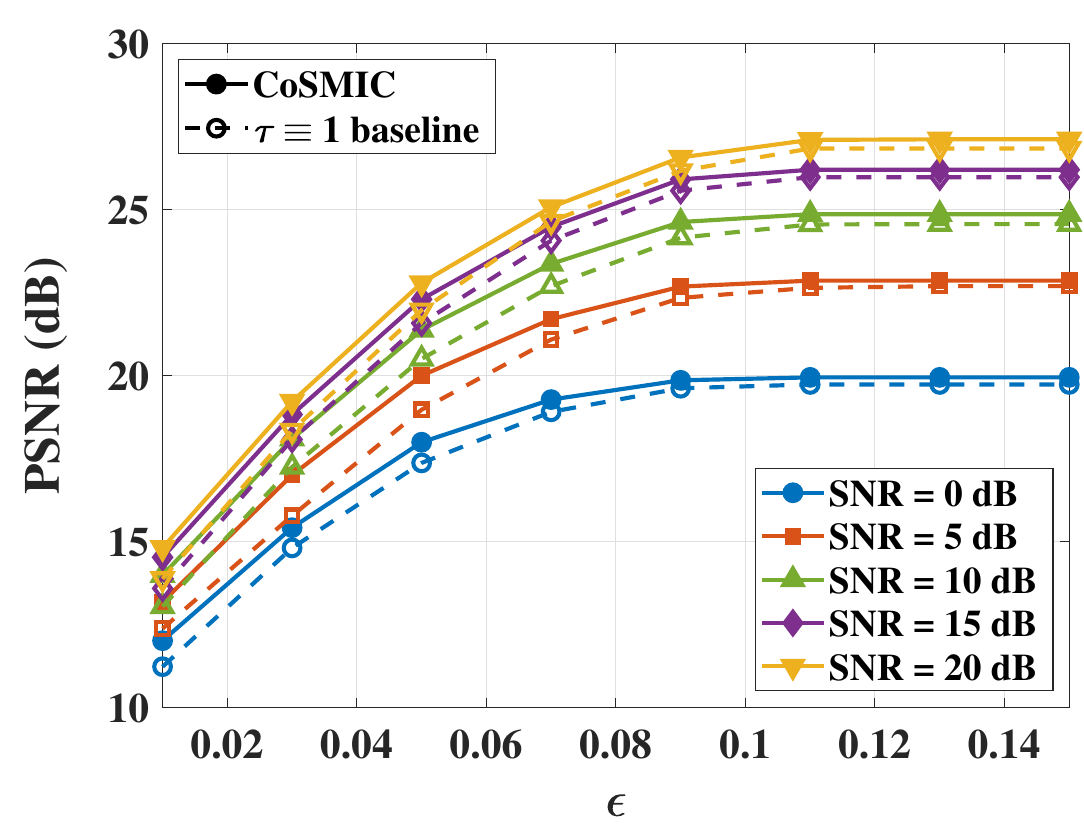}\label{bob_psnr_vs_epsilon}}%
\vspace{-0.9em}
\caption{State recovery and resource trade-offs of CoSMIC. (a) Decomposition of the range error at a Bob SNR of $10$~dB. (b) PSNR~(solid lines) and range RMSE~(dashed lines) versus the state slot share at $J=7680$. (c) PSNR and range RMSE versus the channel-use budget $J$ at a slot share of $17.5\%$. (d) Bob's PSNR versus the covertness budget $\epsilon$, where the dashed lines denote the unit-scale baseline with $\tau_{k,j}\equiv1$.}
\label{state_resource}
\vspace{-0.8em}
\end{figure*}

\begin{figure*}[!t]
\centering
\subfigure[]{\includegraphics[width=0.2398\linewidth]{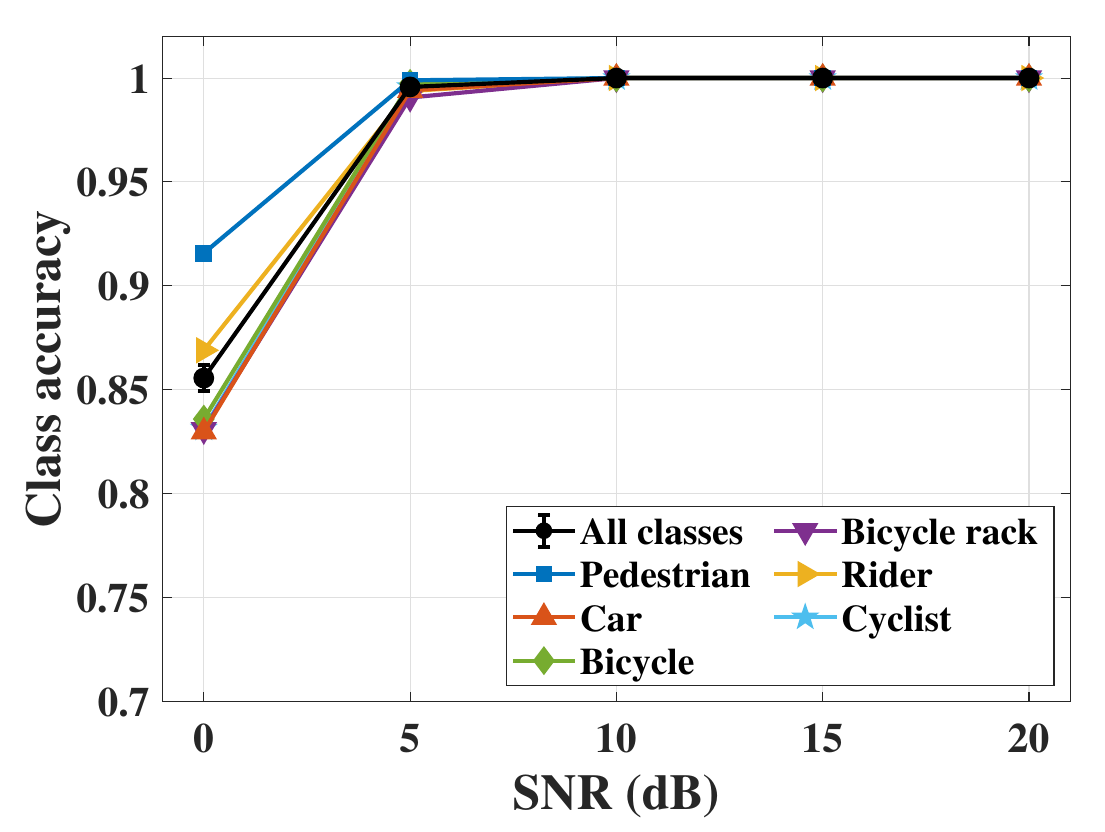}\label{class_accuracy}}\hfill
\subfigure[]{\includegraphics[width=0.2584\linewidth]{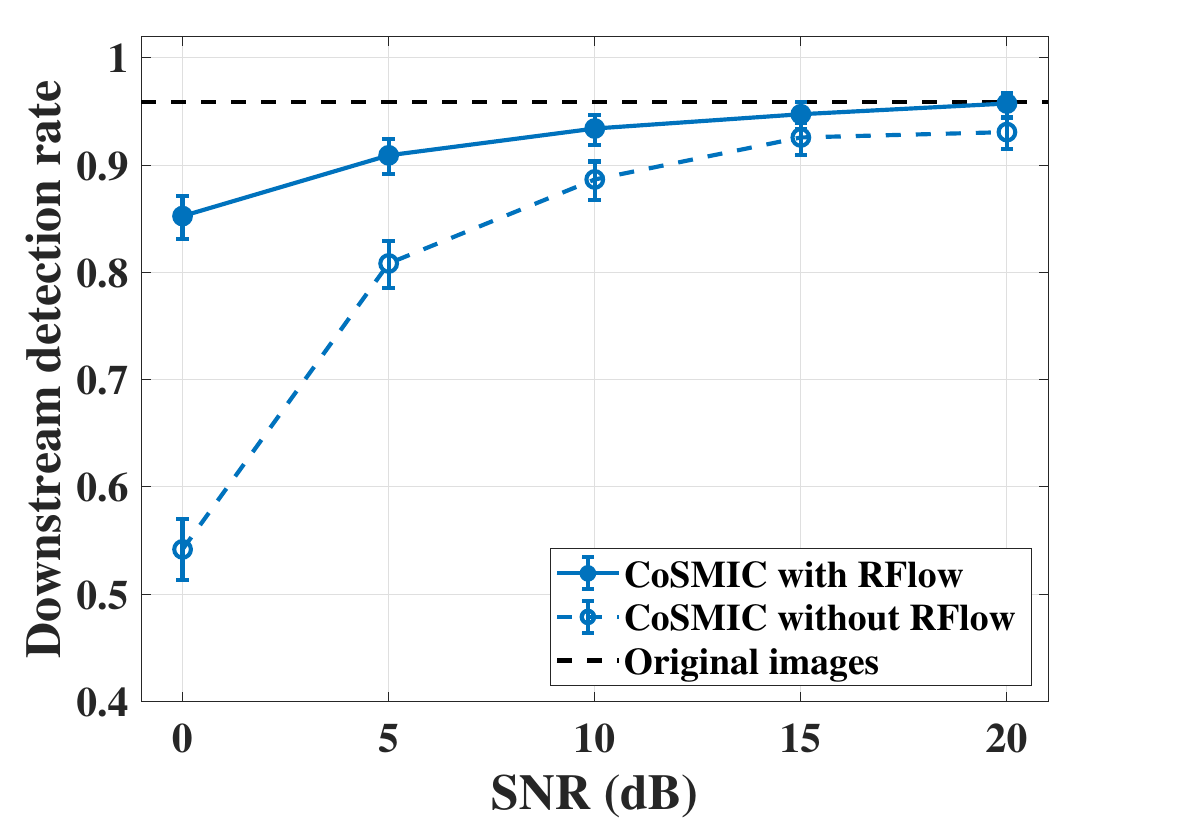}\label{downstream_detection}}\hfill
\subfigure[]{\includegraphics[width=0.2561\linewidth]{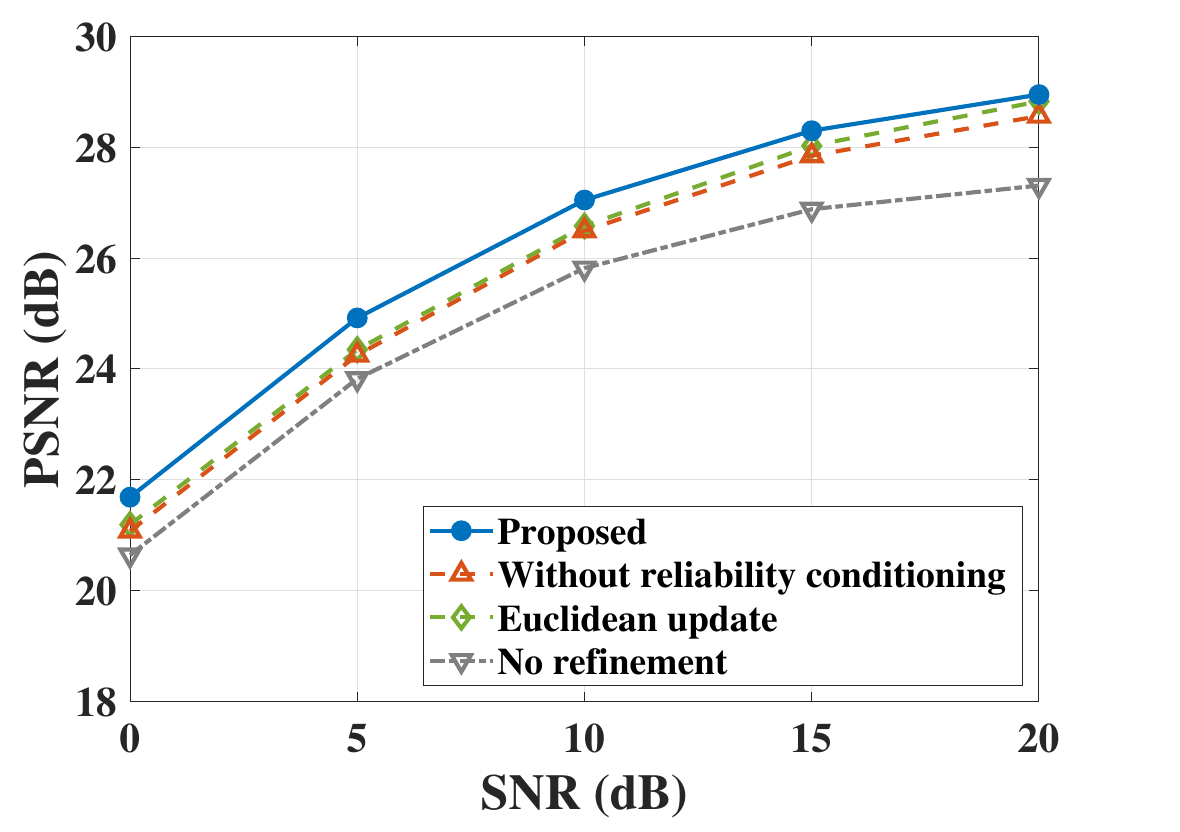}\label{rflow_ablation_psnr}}\hfill
\subfigure[]{\includegraphics[width=0.2353\linewidth]{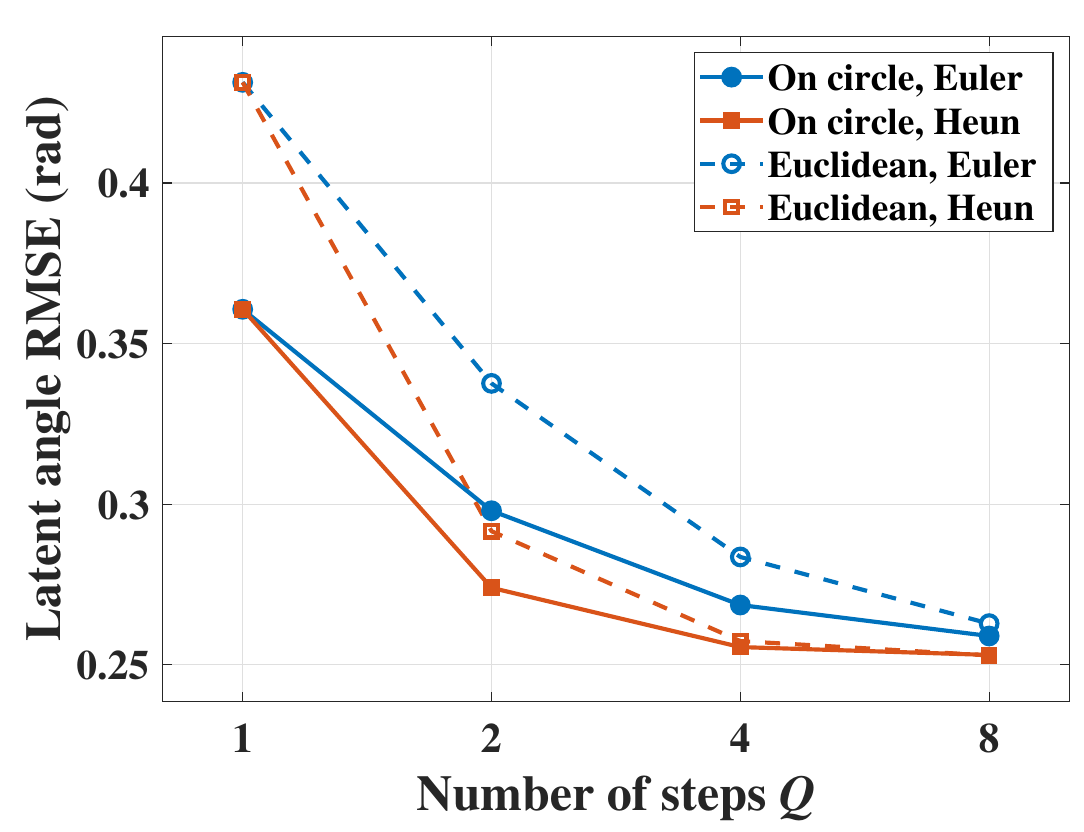}\label{rflow_ablation_latent}}%
\vspace{-0.9em}
\caption{Task-level evaluation and RFlow ablation over the Rayleigh fading channel. (a) Class accuracy. (b) Downstream detection rate. (c) PSNR with every refiner variant at $Q=2$. (d) Latent angle RMSE at an SNR of $10$~dB versus the number of steps $Q$ for the on-circle and the Euclidean updates under the Euler and Heun solvers.}
\label{task_ablation}
\vspace{-0.8em}
\end{figure*}

Beyond the pixel-level fidelity measured by the PSNR, the structural integrity is further evaluated through structural similarity index~(SSIM) in Fig.~\ref{ssim_comparison}.
In the absence of refinement, CoSMIC without RFlow attains an SSIM comparable to the benchmark schemes, and the curves converge at a high SNR.
Once the refinement is activated, CoSMIC with RFlow lifts the SSIM above all schemes and raises it by up to $0.14$ in the low to moderate SNR regimes.
The improvement is more pronounced in the SSIM than in the PSNR, as the velocity field restores the local structural correlations to which the structure-based SSIM is more sensitive.

Fig.~\ref{semantic_examples} presents the reconstructed images and the residual maps of different schemes at SNRs of $0$, $10$, and $20$~dB, where the residual map depicts the pixelwise absolute error against the ground truth and a darker shade marks a larger error.
On the vehicle body, CoSMIC with RFlow produces the lightest residual at every SNR, which reflects the most faithful recovery of the contour and the texture of the car.
Notably, the performance advantage of CoSMIC becomes more pronounced at 0~dB and 10~dB, which confirms that the superiority is sustained under severe channel conditions.

Next, the state component of the semantic message is assessed in Fig.~\ref{radar_rmse} through the RMSE of the recovered range, velocity, and angle against SIMAC, where the transmitted states are the annotated states of the dataset and the RMSE measures the transmission error over the covert link.
The range and velocity RMSEs of CoSMIC remain low and stable across the entire SNR range, remaining an order of magnitude below the corresponding metrics of SIMAC.
Although initially elevated at a low SNR, the angle RMSE of CoSMIC decreases steeply with the SNR, ultimately falling below the range and velocity RMSEs at a high SNR.
The advantage originates from the state coding, since CoSMIC repeats each entry over a reserved group of pairs and recovers the entry through the circular mean.
SIMAC instead carries the states in learned complex symbols decoded by a learned network.
Fig.~\ref{state_error_decomposition} further decomposes the range error of CoSMIC at a Bob SNR of $10$~dB against the radar SNR at Alice into the sensing error of Alice against the ground truth, the transmission error of Bob against the estimate of Alice, and the end-to-end error of Bob against the ground truth.
The transmission error stays constant from a radar SNR of $-35$~dB upward, and the end-to-end error converges to the transmission error from $-30$~dB as the sensing error falls with the radar SNR.
Above a radar SNR of $0$~dB, the covert link determines the state error at Bob, and the RMSE in Fig.~\ref{radar_rmse} therefore characterizes the end-to-end sensing accuracy.
\looseness=-1

Fig.~\ref{slot_share_tradeoff} examines the effect of the slot allocation at $J=7680$, where the share of the state slots is varied from $0$ to $17.5\%$ and the variant without slots allocates all pairs to the image semantics with the states inferred from the decoder features.
The PSNR differs by $0.64$~dB at an SNR of $10$~dB between the shares of $0$ and $17.5\%$, since the state slots occupy at most $1344$ of the $7680$ pairs.
In contrast, the range RMSE at $20$~dB drops from $4.37$~m without slots to $0.12$~m, since the target range is weakly observable in a single image and the averaging gain of the slot decoding grows with the group size.
The above observation indicates that the dedicated slots attain an accurate state recovery at a marginal share of the channel uses, which justifies the explicit transmission of the states.

Fig.~\ref{budget_tradeoff} evaluates CoSMIC under different communication budgets, where the number of pairs per frame is reduced from $J=7680$ to $3840$ and $1920$ at the slot share of $17.5\%$ and the bandwidth ratio decreases from $1/12$ to $1/24$ and $1/48$.
The PSNR drops from $26.42$ to $23.76$ and $22.22$~dB at an SNR of $10$~dB, since the image semantics are spread over fewer pairs at a smaller $J$.
In contrast, the range RMSE at $20$~dB rises from $0.12$ to $0.21$ and $0.27$~m, which indicates that the budget mainly governs the image fidelity whereas the state recovery remains accurate under a reduced budget.

\begin{figure}[t]
  \centering
  \includegraphics[width=0.8\linewidth]{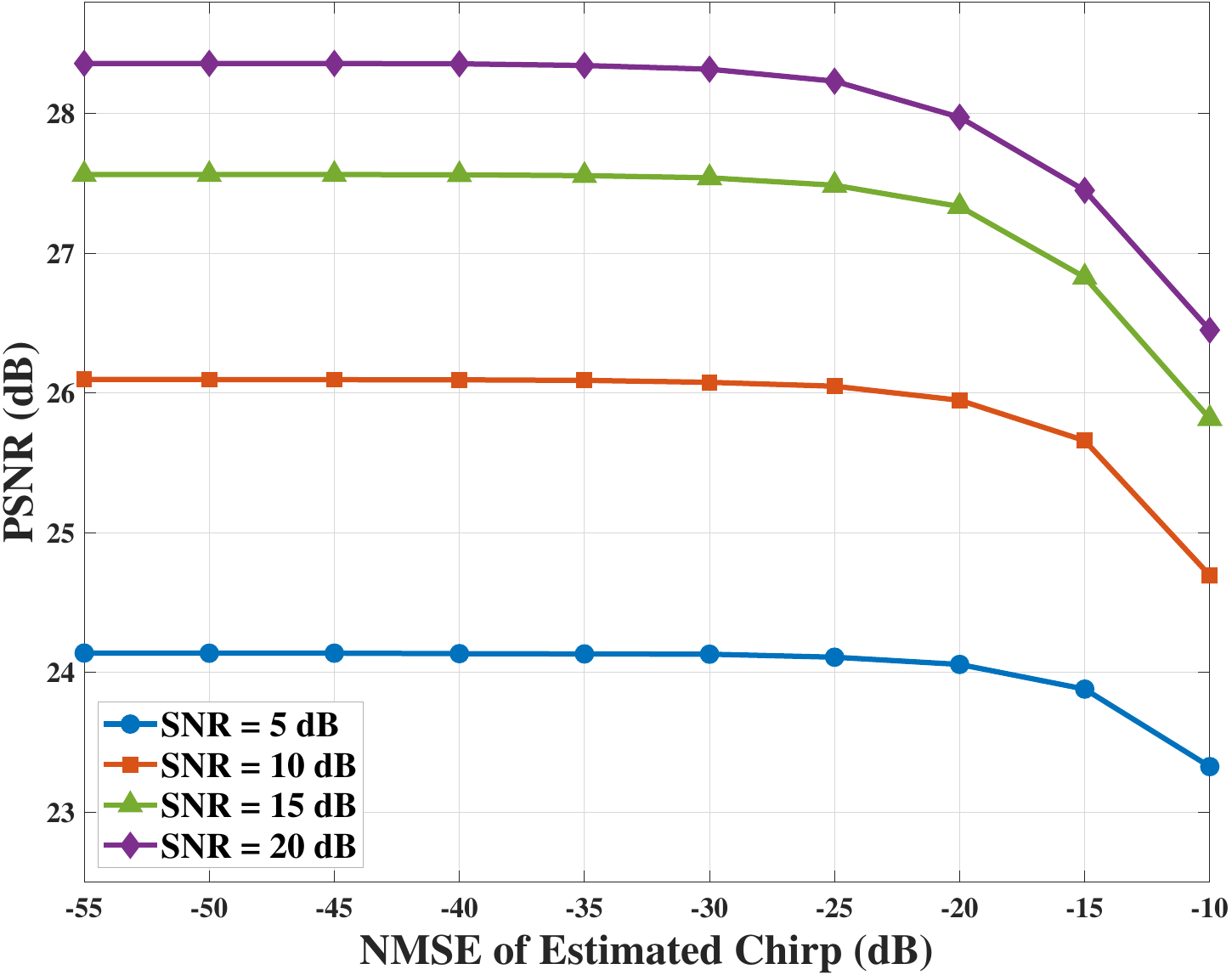}
\vspace{-0.4em}
  \caption{Bob's PSNR versus the NMSE of the estimated chirp basis under different SNRs.}
  \label{bob_psnr_vs_basis_nmse_db}
\vspace{-0.6em}
\end{figure}

Fig.~\ref{bob_psnr_vs_epsilon} examines the tradeoff between the SemCom performance and the covertness through Bob's PSNR versus the covertness budget $\epsilon$.
All curves are obtained with the model trained at $\epsilon=0.10$ under the budget set at inference.
Under a tight covertness constraint, the bound in Theorem~\ref{thm:covertness} forces the pair scales toward unity, which suppresses the deviation of the transmitted field from the Gaussian reference and reduces the semantic information embedded by the rotation coding, thereby lowering the PSNR.
As the budget loosens, the PSNR rises steeply and saturates for $\epsilon$ above $0.1$, where the recovery becomes limited by the channel noise instead of the covertness budget, and the saturated level increases with the SNR.
Fig.~\ref{bob_psnr_vs_epsilon} further includes a separately trained unit-scale baseline, which fixes all pair scales to $\tau_{k,j}\equiv1$ in both training and inference.
CoSMIC outperforms the baseline at each budget and SNR with a PSNR gain of up to $1.2$~dB, which quantifies the contribution of the scale modulation to the semantic reconstruction.
\looseness=-1

Fig.~\ref{class_accuracy} and~\ref{downstream_detection} assess the recovered semantics at the task level, where the target class carried by the reserved slots is decoded at Bob over $12139$ test targets of ten classes and a COCO-pretrained Faster R-CNN detector is applied to the reconstructed images of $1200$ test frames.
The class accuracy rises from $0.856$ at $0$~dB to $0.996$ at $5$~dB and reaches $1.000$ from $10$~dB onward.
With the detection rate of $0.959$ on the original images as the reference, the detection rate of CoSMIC with RFlow reaches $0.853$ at $0$~dB and stays above $0.93$ from $10$~dB onward, whereas the rate without RFlow is $0.542$ at $0$~dB.
The results indicate that the recovered semantics directly serve the downstream perception task.
\looseness=-1

Fig.~\ref{rflow_ablation_psnr} and~\ref{rflow_ablation_latent} examine the contributions of the reliability conditioning and the on-circle update of the RFlow refiner, where every refiner variant in Fig.~\ref{rflow_ablation_psnr} runs $Q=2$ steps.
The variant without reliability conditioning is retrained with zero reliability inputs, and the variant with the Euclidean update advances every pair in the plane without the constraint to the unit circle.
The proposed refiner raises the PSNR of the decoder without refinement by $1.0$ to $1.6$~dB over the SNR range, where the reliability conditioning contributes $0.4$ to $0.7$~dB and the on-circle update contributes $0.1$ to $0.6$~dB over the Euclidean update.
In Fig.~\ref{rflow_ablation_latent}, the latent angle RMSE at an SNR of $10$~dB decreases monotonically with $Q$ for the on-circle update under both solvers, and the deployed setting $Q=2$ attains the PSNR of the larger $Q$ with two velocity-field evaluations.
\looseness=-1

Fig.~\ref{bob_psnr_vs_basis_nmse_db} evaluates the robustness of the SemCom performance against the chirp estimation error at Bob, where the NMSE of the estimated chirp basis quantifies the estimation accuracy.
For an NMSE below $-20$~dB, the PSNR remains stable across all SNRs, which indicates the resilience of the receiver to a moderate chirp estimation error.
As the NMSE rises above $-20$~dB, the PSNR degrades by up to $2$~dB, since the residual de-chirp error perturbs the phase extraction that feeds the latent recovery.
The degradation is more pronounced at a high SNR owing to the dominant chirp error, whereas channel noise conceals the impairment at a low SNR.

\begin{table}[t]
\centering
\caption{Per-frame latency comparison of CoSMIC and baselines.}
\label{tab:latency_comparison}
\vspace{-0.4em}
\begin{tabular}{lcc}
\hline
\textbf{Method} & \textbf{Transmitter (ms)} & \textbf{Receiver (ms)} \\
\hline
CoSMIC without RFlow & $8.380 \pm 0.316$ & $6.880 \pm 0.032$ \\
CoSMIC with RFlow   & $8.380 \pm 0.316$ & $9.435 \pm 0.431$ \\
HPQ~\cite{yilmaz2024high}             & -- & $13.749 \pm 0.281$ \\
CDDM~\cite{wu2024cddm}            & -- & $22.678 \pm 0.190$ \\
\hline
\end{tabular}
\vspace{-0.8em}
\end{table}

Table~\ref{tab:latency_comparison} compares the per-frame latency of CoSMIC and the diffusion-based benchmarks at the deployed configurations of Fig.~\ref{performance_comparison}, where each scheme runs the sampling budget of its deployed configuration at a batch size of one.
Integrating the RFlow refiner with $Q=2$ steps introduces a minor overhead of approximately 2.5 ms, which keeps the overall receiver latency within $10$~ms.
Compared with HPQ and CDDM, the proposed CoSMIC with RFlow reduces the receiver latency by $31\%$ and $58\%$, respectively.
The receiver latency covers the processing from the received frame to the decoded message, and the sensing result of a frame is delivered in the following frame.
The delay from the echo acquisition to the decoded message therefore comprises two frame durations of $2.56$~ms and the transmitter and receiver latencies, which stays below $23$~ms within the $100$~ms interval of a $10$~Hz perception frame rate.
\looseness=-1

\begin{figure*}[b]
\vspace{-0.3em}
\hrulefill
\begin{equation}\label{eq:kl_gaussian}
\begin{aligned}
D\left(P_{\mathbf{b},1,k}\,\middle\|\,P_{\mathbf{b},0,k}\right)
=\frac{1}{2}\sum_{j=1}^{J}\left[\mathrm{tr}\left(\tau_{k,j}\mathbf{I}_2\right)-2-\ln\det\left(\tau_{k,j}\mathbf{I}_2\right)\right]
=\sum_{j=1}^{J}\left(\tau_{k,j}-\ln\tau_{k,j}-1\right)=\sum_{j=1}^{J}\chi(\tau_{k,j}),      
\end{aligned} 
\end{equation}

\begin{equation}\label{eq:multi_target_response}
\begin{aligned}
R_{k,s}^{(i)}(d_{k,\ell},\nu_{k,\ell},\theta_{k,\ell})=\alpha_{k,\ell}UMN
+\sum_{r\ne\ell}\alpha_{k,r}S(\theta_{k,\ell},\theta_{k,r})A_k^{(i)}(\Delta_{\ell r},\Gamma_{\ell r}),
\end{aligned}
\end{equation}

\begin{equation}\label{eq:interference_bound}
\begin{aligned}
\left|I_{k,\ell}^{(1)}\right| = \left|\sum_{r\ne\ell}\alpha_{k,r}S(\theta_{k,\ell},\theta_{k,r})A_k^{(1)}(\Delta_{\ell r},\Gamma_{\ell r})\right|
\le \sum_{r\ne\ell}|\alpha_{k,r}|\,|S(\theta_{k,\ell},\theta_{k,r})|\,B_{A,k}\!\left(\Delta_{\ell r},\Gamma_{\ell r};\tfrac{\eta}{R_k^{\mathrm{I}}}\right).
\end{aligned}
\end{equation}

\begin{align} \label{eq:cross_bounds}
\left|A_k^{(1)}(\Delta_{\ell r},\Gamma_{\ell r})\right| &\le B_{A,k}(\Delta_{\ell r},\Gamma_{\ell r};\eta'), \quad \left|A_k^{(1)}(\Delta_{\ell r},\Gamma_{\ell r})-A_k^{(0)}(\Delta_{\ell r},\Gamma_{\ell r})\right| \le B_{E,k}(\Delta_{\ell r},\Gamma_{\ell r};\eta'),
\end{align}

\vspace{-0.7em}
\end{figure*}



\section{Conclusion}\label{sec:conclusion}
\vspace{-0.25em}


In this paper, CoSMIC was proposed as a covert semantic ISAC system that embeds sensing-derived semantics into a dual-functional chirp waveform through Gaussian pair rotation coding.
A closed-form multi-frame covertness constraint was derived, and the budget was enforced through the covert projection.
Additionally, the preserved radar response was characterized through the mainlobe invariance and a lower bound on the output SINR.
Furthermore, a reliability-guided RFlow refiner was developed for latent recovery, and the simulation results demonstrated improved semantic fidelity with guaranteed covertness and preserved sensing accuracy.

\vspace{-1em}
\begin{appendices}

\section{Proof of Theorem~\ref{thm:covertness}}\label{app:proof_theorem1}

The covertness of the transmission over the window is governed by the divergence $D(P_{1}^{(K)}\|P_{0}^{(K)})$ between the hypotheses $H_1$ and $H_0$ of Willie.
Let $\mathcal{S}$ collect the scene process and the radar noise over the window, where $\mathcal{S}$ follows an identical distribution under both hypotheses and is independent of the reference pairs.
Since the semantic message of each frame is generated from the sensing output of the previous frame through~\eqref{eq:semantic_payload}, the rotation matrix $\mathbf{A}_k$ is determined by $\mathcal{S}$ and the symbol vectors of the previous frames, whereas the reference vector $\mathbf{q}_k$ is independent of both owing to the fresh reference pairs.
Conditioned on $\mathcal{S}$ and the previous symbol vectors, the symbol vector in~\eqref{eq:frontend_fields} follows $\mathcal{N}(\mathbf{0},\mathbf{I}_{MN})$ under $H_0$ and $\mathcal{N}(\mathbf{0},\mathbf{A}_k\mathbf{A}_k^{\mathsf{T}})$ under $H_1$.
Direct computation from~\eqref{eq:rotation_matrix} yields $\mathbf{A}_{k,j}\mathbf{A}_{k,j}^{\mathsf{T}}=\tau_{k,j}\mathbf{I}_2$, and therefore $\mathbf{A}_k\mathbf{A}_k^{\mathsf{T}}=\mathrm{blkdiag}(\tau_{k,1}\mathbf{I}_2,\ldots,\tau_{k,J}\mathbf{I}_2)$.
The conditional KL divergence between the zero-mean Gaussian distributions under $H_1$ and $H_0$ thus decomposes over the diagonal blocks as~\eqref{eq:kl_gaussian}, where $P_{\mathbf{b},i,k}$ denotes the conditional distribution of $\mathbf{b}_k^{(i)}$ given $\mathcal{S}$ and the previous symbol vectors.

The divergence in~\eqref{eq:kl_gaussian} pertains to the symbol vector, whereas Willie observes the transmitted waveform, and the two domains are connected by the symbol-to-waveform chain.
The chain comprises the shaping $\mathbf{u}_k^{(i)}=\mathbf{G}_{\rho}\mathbf{b}_k^{(i)}$ with invertible $\mathbf{G}_{\rho}$, the elementwise mapping $u\mapsto \pi\tanh(\alpha_{\phi}u)$ that is a strict bijection from $\mathbb{R}$ onto $(-\pi,\pi)$, and the per-sample modulation $\phi\mapsto p_n e^{j\phi}$ that is one-to-one on $(-\pi,\pi)$.
The KL divergence is invariant under a bijective transformation, and the composition of the three bijections preserves the divergence along the entire chain.
The waveform distributions therefore satisfy
\begin{equation}\label{eq:kl_waveform}
{D\left(P_{\mathbf{x},1,k}\,\middle\|\,P_{\mathbf{x},0,k}\right)}
= D(P_{\mathbf{b},1,k} \| P_{\mathbf{b},0,k})
= \sum_{j=1}^{J}\chi(\tau_{k,j}),
\end{equation}
where $P_{\mathbf{x},i,k}$ denotes the conditional distribution of $\mathbf{x}_k^{(i)}$ given $\mathcal{S}$ and the previous symbol vectors.

Willie forms the binary decision from the observation over the window, which is induced from the transmitted waveforms of the $K$ frames through the common channel transition laws $W_k(\cdot\mid\mathbf{x})$ in~\eqref{eq:frame_marginal} under both hypotheses.
The data processing inequality for the KL divergence then states
\begin{equation}\label{eq:dpi}
D\left(P_{1}^{(K)}\,\middle\|\,P_{0}^{(K)}\right)\le D\left(P_{\mathbf{x},1}^{(K)}\,\middle\|\,P_{\mathbf{x},0}^{(K)}\right),
\end{equation}
where $P_{\mathbf{x},i}^{(K)}$ denotes the joint distribution of $\mathcal{S}$ and the transmitted waveforms of the $K$ frames under $H_i$.
Since $\mathcal{S}$ follows an identical distribution under both hypotheses and the previous waveforms are a bijective function of the previous symbol vectors, the chain rule of the KL divergence decomposes the right-hand side over the frames as
\begin{equation}\label{eq:kl_chain}
D\left(P_{\mathbf{x},1}^{(K)}\,\middle\|\,P_{\mathbf{x},0}^{(K)}\right)=\sum_{k=1}^{K}\mathbb{E}\left[D\left(P_{\mathbf{x},1,k}\,\middle\|\,P_{\mathbf{x},0,k}\right)\right],
\end{equation}
where the expectation is taken over $\mathcal{S}$ and the previous symbol vectors under $H_1$.
The combination of~\eqref{eq:dpi} and~\eqref{eq:kl_chain} with~\eqref{eq:kl_waveform} establishes~\eqref{eq:kl_bound}.
Under the budget~\eqref{eq:covert_budget}, the right-hand side of~\eqref{eq:kl_bound} is bounded by $2\epsilon^{2}$, and substitution into the covertness bound~\eqref{eq:pinsker_covertness} lower bounds the optimal detection error probability of Willie over the window by $1-\epsilon$, which completes the proof of Theorem~\ref{thm:covertness}.


\section{Proof of Theorem~\ref{thm:radar}}\label{app:proof_theorem2}

Since $x_{k,m,n}^{(i)}=p_n e^{j\phi_{k,m,n}^{(i)}}$ with $|x_{k,m,n}^{(i)}|=1$, the ambiguity function \eqref{eq:ambiguity} at zero range lag reduces to
\begin{equation}\label{eq:ambiguity_zero}
A_k^{(i)}(0,\Gamma)=\sum_{m=0}^{M-1}\sum_{n=0}^{N-1}\left|x_{k,m,n}^{(i)}\right|^{2}e^{-j2\pi\Gamma m/M}=N D_M(\Gamma),
\end{equation}
which is identical under $H_0$ and $H_1$.

For the single-target case with $L_k=1$, substitution of~\eqref{eq:ambiguity_zero} into the signal component of~\eqref{eq:matched_mode} provides
\begin{equation}\label{eq:single_target_response}
R_{k,s}^{(i)}(d_{k,1},\nu,\theta)=\alpha_{k,1} S(\theta,\theta_{k,1})N D_M(\nu-\nu_{k,1}).
\end{equation}
Evaluating~\eqref{eq:single_target_response} at the true bin $(d_{k,1},\nu_{k,1},\theta_{k,1})$ with $S(\theta_{k,1},\theta_{k,1})=U$ and $D_M(0)=M$ yields $R_{k,s}^{(i)}(d_{k,1},\nu_{k,1},\theta_{k,1})=\alpha_{k,1} UMN$ for both modes, which establishes~\eqref{eq:mainlobe_invariance}.
Since the filtered noise has power $\sigma_r^{2}UMN$, the output SNR of $H_0$ and $H_1$ equals
\begin{equation}\label{eq:single_snr}
\mathrm{SNR}_k^{(i)}=\frac{|\alpha_{k,1} UMN|^{2}}{\sigma_r^{2}UMN}=\frac{|\alpha_{k,1}|^{2}UMN}{\sigma_r^{2}}, i\in\{0,1\}.
\end{equation}

For the multi-target case with $L_k\ge 2$, the signal component of~\eqref{eq:matched_mode} at the true bin of the $\ell$-th target separates into a self term and the cross terms of the remaining targets, which is given by~\eqref{eq:multi_target_response}, where the self term $\alpha_{k,\ell}UMN$ is independent of the transmit mode by \eqref{eq:ambiguity_zero}.
According to the concentration property established in Section~\ref{subsec:radar_analysis}, for any $\eta'\in(0,1)$, each cross term is bounded with probability at least $1-\eta'$ by~\eqref{eq:cross_bounds}, where the envelopes follow from~\eqref{eq:envelopes} after combining the mean and the deviation bounds through the triangle inequality.

Setting $\eta'=\eta/R_k^{\mathrm{I}}$ in the second inequality of~\eqref{eq:cross_bounds} and applying the union bound over the $R_k^{\mathrm{I}}=L_k-1$ interferers, the bound holds for all $r\ne\ell$ simultaneously with probability at least $1-\eta$.
Subtracting~\eqref{eq:multi_target_response} for both modes and invoking the triangle inequality yields~\eqref{eq:response_dev}.
Similarly, setting $\eta'=\eta/R_k^{\mathrm{I}}$ in the first inequality of~\eqref{eq:cross_bounds} and applying the union bound, the interference of the $\ell$-th target under $H_1$ is bounded with probability at least $1-\eta$ by~\eqref{eq:interference_bound}.
Given the self-signal power $|\alpha_{k,\ell}UMN|^{2}$ and the noise power $\sigma_r^{2}UMN$, the output SINR satisfies~\eqref{eq:sinr_bound}.
This completes the proof of Theorem~\ref{thm:radar}.

\end{appendices}

\bibliographystyle{IEEEtran}
\bibliography{reference}

\end{document}